\documentclass[5p]{elsarticle}
\usepackage{amssymb}
\usepackage[tbtags]{amsmath}
\usepackage{graphicx}
\usepackage{natbib}
\usepackage{geometry}
\usepackage{xcolor} 
\usepackage{algorithm}
\usepackage[noend]{algpseudocode}
\usepackage{mathtools}

\newcommand{\x}{\mathbf{x}}
\newcommand{\hth}{\hat{\theta}}
\newcommand{\hnu}{\hat{\nu}}

\begin{document}

\title{Parallel inverse-problem solver for time-domain optical
  tomography\\ with perfect parallel scaling}

\author[1,2]{E. L. Gaggioli\corref{cor1}}
\ead{egaggioli@iafe.uba.ar}
\author[3]{and O. P. Bruno}

\address[1]{Instituto de Astronom\'{\i}a y F\'{\i}sica del Espacio
(IAFE, CONICET--UBA), Casilla de Correo 67 - Suc. 28 (C1428ZAA), Buenos Aires, Argentina.}
\address[2]{Departamento de F\'{\i}sica, FCEyN, Universidad de Buenos Aires.}
\address[3]{Computing and Mathematical Sciences, Caltech, Pasadena, CA 91125, USA.}

\date{\today}


\begin{abstract}
  This paper presents an efficient {\em parallel} radiative
  transfer-based inverse-problem solver for time-domain optical
  tomography. The radiative transfer equation provides a physically
  accurate model for the transport of photons in biological tissue,
  but the high computational cost associated with its solution has
  hindered its use in time-domain optical-tomography and other
  areas. In this paper this problem is tackled by means of a number of
  computational and modeling innovations, including 1) A spatial
  parallel-decomposition strategy with \textit{perfect parallel
    scaling} for the forward and inverse problems of optical
  tomography on parallel computer systems; and, 2) A Multiple
  Staggered Source method (MSS) that solves the inverse transport
  problem at a computational cost that is {\em independent of the
    number of sources employed}, and which significantly accelerates
  the reconstruction of the optical parameters: a six-fold MSS
  acceleration factor is demonstrated in this paper. Finally, this
  contribution presents 3) An intuitive derivation of the
  adjoint-based formulation for evaluation of functional gradients,
  including the highly-relevant general Fresnel boundary
  conditions---thus, in particular, generalizing results previously
  available for vacuum boundary conditions. Solutions of large and
  realistic 2D inverse problems are presented in this paper, which
  were produced on a 256-core computer system. The combined
  parallel/MSS acceleration approach reduced the required computing
  times by several orders of magnitude, from months to a few hours.
\end{abstract}
\maketitle
\section{Introduction}
\label{sec:introduction}

This paper presents an efficient parallel inverse-problem solver,
based on the radiative transfer equation (RTE), for time-domain
optical tomography.  As is well known, the RTE provides a physically
accurate model for the transport of photons in biological tissue
\cite{Klose2009,Arridge2009}, but its applicability in the context of
inverse problems has been hindered by the high computational cost
required for its solution.  In this paper this difficulty is addressed
by means of a combination of three main strategies, namely, (i)~Use of
the spectral FC-DOM approach~\cite{Gaggioli2019} for the solution of
the RTE equations; (ii)~An effective parallel implementation of the
FC-DOM method based on a spatial domain-decomposition strategy; and
(iii)~A Multiple Staggered Source (MSS) setup, which utilizes certain
combinations of sources operating simultaneously instead of the
sequences of single sources used in previous approaches. When
implemented on a 256-core computer cluster the overall parallel/MSS
approach results in computing-time acceleration by several orders of
magnitude---thus making it possible to solve large 2D inverse
problems, such as the problem of imaging within a neck model
considered in Section~\ref{sec:inverseres}.

The proposed direct parallel solvers provide a number of
advantages. In contrast to other algorithms, for example, the approach
presented in this contribution is highly efficient independently of
the number of sources employed (cf.~\cite{Hielscher2004}).
Alternative parallel strategies based on GPU
architectures~\cite{Doulgerakis2017} might be appropriate for optical
tomography based on the diffusion approximation. But, unfortunately,
the diffusion approximation is not accurate for many applications,
and, in view of the high storage required by the RTE-based time-domain
inverse problem, a GPU-based RTE solver may not be viable. In the
review~\cite{Coelho2014} of parallelization strategies for this
problem, for strong scaling tests, all parallelization strategies show
an efficiency scaling significantly below the
ideal. Ref.~\cite{Colomer2013} presents the only parallel strategy
known to the authors which reports a perfect parallel efficiency, but
the applicability of the method is restricted by a fundamental
limitation to non-absorbing and non-scattering media.  In contrast,
the parallel algorithm presented in this paper enjoys perfect parallel
efficiency for general problems, including arbitrarily prescribed
scattering and absorption, and finds no restrictions regarding the
transport regime, the number of sources, or the number of discrete
ordinates employed.  A parallel efficiency of 136.7\% with up to 256
cores for the benchmark presented in this work has been obtained by
means of the proposed parallelization for the FC-DOM algorithm (see
sec.~\ref{subsec:FCGram} Figure~\ref{fig:scala} and its caption). For
comparison, reference \cite[p. 153]{Fujii2014} reports a computing
time of 44.3 hours for a single 2D forward time solution of the RTE,
which, running on 64 cores, the algorithms presented in this paper
obtain in less than thirty minutes.

As mentioned above, an additional significant reduction in the
computing time required for the solution of the inverse problem is
achieved by exploiting the proposed MSS method---which, combining
multiple sources in each RTE solution, reduces the
number of time-domain direct and adjoint RTE solutions required. MSS
acceleration by a factor of six is demonstrated in this paper, without
any degradation in the accuracy of the inverse problem reconstruction,
relative to the time required by the Transport Sweep
method~\cite{Prieto2017,Dorn} (TS) ubiquitously encountered in optical
tomography.

The rest of this paper is organized as follows. Section
\ref{sec:MathIntro} presents the classical optical-tomography forward
problem. Section~\ref{sec:inversep} then presents a derivation of the
adjoint gradient formalism under the highly relevant general Fresnel
boundary conditions (thus generalizing results previously available
for vacuum boundary conditions) which, in particular, incorporates a
class of \textit{generalized sources} inherent in the MSS method in
addition to the classical source types inherent in the TS method. A
direct comparison, presented in Section~\ref{sec:gradc}, between
gradients for the objective function obtained by finite differences
and the adjoint method demonstrate the validity and accuracy of the
proposed adjoint approach. Section~\ref{sec:nmethods}, in turn,
presents the numerical methods employed for the numerical solution of
the RTE, as well as the spatial domain decomposition strategy utilized
for their efficient parallelization. Section~\ref{sec:forwsim}
demonstrates the high-order convergence of the overall RTE parallel
solver for smooth solutions, and it presents a demonstration for a
photon density of a type used in laboratory settings.  Section
\ref{sec:inverseres} presents results of a number of reconstructions
obtained by the proposed methods, including large 2D inverse problems,
such as the problem of imaging within a head model and an MRI-based
neck model.

\section{Preliminaries}
\label{sec:MathIntro}

Defining the transport operator
\begin{equation}
\begin{split}
\mathcal{T}[u]&=\frac{1}
{c}\frac{\partial u(\x, \theta,t)}{\partial t} + \hth \cdot \nabla u(\x,\theta,t)+a(\x)u(\x,\theta,t) \\
&+b(\x)\left[u(\x,\theta,t) - \int_{S^1}\eta(\hth\cdot\hth ') u(\x,\theta',t) d\theta'\right],
\end{split}
\label{eq:Toper}
\end{equation}
we consider the initial and boundary-value radiative transfer problem
\begin{equation}
\begin{split}
\begin{aligned}
&\mathcal{T}\big[u\big]=0, \;  (\x,\theta)  \in \Omega\times [0,2\pi)\\
&u(\x,\theta,t=0)=0, \;  (\x,\theta)  \in \Omega\times [0,2\pi) \\
&u(\x,\theta,t)=f(\hth \cdot \hnu) u(\x,\theta_r,t) + q(\x,\theta,t) , \; (\x,\theta) \in \Gamma_-
\end{aligned}
\end{split}
\label{eq:RTE}
\end{equation}
for the amount of energy $u=u(\x,\theta,t)$ irradiated at a point
$\x$, propagating with direction $\hth \in S^1$, at a time $t$, over
the two-dimensional spatial domain $\Omega$ and for $0\leq t\leq T$
($T>0$), where $\hth = (\cos(\theta),\sin(\theta))$
($0\leq \theta\leq 2\pi$), $\hth' = (\cos(\theta'),\sin(\theta'))$
($0\leq \theta'\leq 2\pi$), and where $S^1$, $c$, $a(\x)$ and $b(\x)$
denote the unit circle, the average speed of light in the medium, and
the absorption and scattering coefficients, respectively. Further,
$\hnu$ denotes the outward unit vector normal to the boundary
$\partial \Omega$ of the domain $\Omega$,
$\Gamma_{\pm}=\{(\x,\theta) \in \partial \Omega \times [0,2\pi), \pm
\hat \nu \cdot \hth >0 \}$---with subindex $-$ (resp. $+$) denoting
the set of incoming (resp. outgoing) directions---and $q(\x,\theta,t)$
denotes a given source function, which can be used to model
illuminating lasers that inject radiation over some portion of
$\Gamma_-$. Additionally, the function
$f(\hth \cdot \hnu)=f(\cos(\theta-\theta_{\nu}))$ denotes the Fresnel
coefficient, according to which radiation is reflected into $\Omega$
and transmitted away from $\Omega$ in the directions
$\hth_r=(\cos(\theta_r),\sin(\theta_r))$ and
$\hat \theta_\mathrm{tr} = \hat\alpha(\hat \theta)$ respectively,
where, in accordance with Snells' law for the refractive indexes
$n_{\Omega}$ and $n_0$ within and outside $\Omega$,
$\theta_r=R(\theta)=2\theta_{\nu}-\theta+\pi$ (mod $2\pi$) and
$ \hat \alpha(\hat \theta)$ denote a reflection function and the
transmission angle, respectively. And, finally
\begin{equation}
\eta(\hth\cdot\hth')=\frac{1}{2\pi} 
\frac{1-g^2}{(1+g^2-2g \, \hth \cdot \hth')^{3/2}}
\label{eq:Henyey-Greeinstein}
\end{equation}
denotes the widely used Henyey--Greeinstein phase
function~\cite{Henyey1941} which models the angle dependent
redistribution probability after a photon collision event. The index
$g \in [-1,1]$ characterizes the scattering anisotropy: pure forward
scattering (resp. pure backward scattering), wherein all photons
emerge in the forward (resp. backward) direction after a collision
event, corresponds to a value of $g=1$ (resp. $g=-1$). Isotropic
scattering, in turn, where a photon emerges with equal probability in
any given direction after collision corresponds to the value
$g=0$. The total density of photons at a particular point $\x$ at a
time $t$, which is considered in particular in the context of the
photon diffusion approximation, is quantified by the scalar flux
\begin{equation}
  \Phi(\x,t)=\int_{S^1} u(\x,\theta,t) d\theta.
\label{eq:photondensity}
\end{equation}

To conclude this section we mention the window function~\cite{Bruno2014}
\begin{equation}
\begin{aligned}
\begin{split}
w(v)=
\begin{cases}
  1 &\text{for} \, \, v = 0, \\
       \exp{\left(\frac{2e^{-1/|v|}}{|v|-1}\right)} & \text{for} \, \, 0 < |v| < 1, \\
       0 &\text{for} \, \, |v| \geq 1.
     \end{cases}
\end{split}
\end{aligned}
\label{eq:sourcelabwind}
\end{equation}
of the real variable $v$, which vanishes for $|v|\geq 1$ and smoothly
transitions to one in the interval $-1 < v < 1$. This function is used
in multiple roles in what follows---including modeling of both the
temporal profile and the collimated irradiation of the laser beam, as
well as the spatial dependence of the sensitivity of the detectors.
\section{Inverse problem and adjoint gradient formalism}
\label{sec:inversep}
\subsection{Objective functions}\label{sec:formulat}
This section introduces the general form of the {\em objective
  functions} whose minima yield the solutions of the TS and MSS
inverse transport problems we consider, as well as the {\em adjoint
  formalism} we use for the efficient evaluation of the corresponding
objective-function gradients under general Fresnel boundary
conditions---that is, using a possibly non-vanishing Fresnel
coefficient $f$ in the last line in equation~\eqref{eq:RTE}.  For
$f(\hth \cdot \hnu)\equiv 0$ the gradient expression we obtain
coincides with the well-known result for homogeneous boundary
conditions that are relevant in neutron
transport~\cite{Duderstadt1979} (vacuum boundary conditions) and that
are also often assumed in the context of optical
tomography~\cite{Prieto2017,Dorn,Ren2010}.

The TS and MSS objective functions are based on use of sources of two
different types, both of which can expressed in the form
\begin{equation}
  q = q_i(\x,\theta,t) =\sum_{k=1}^{N_s} s_{k,i}(\x,\theta,t) ,\quad i=1,2,\ldots,N_q
\label{eq:RTEsources}
\end{equation}
for certain integer values of $N_q$ and $N_s$.  Here, using the window
function~\eqref{eq:sourcelabwind}, we define
\begin{equation}
  s_{k,i}(\x,\theta,t) = \exp(-|\x-\x_{k,i}|^2/
  2\sigma_\x^2)w(\delta_{k,i})T(t-\tau_{k,i}),
\label{eq:RTEsources2}
\end{equation}
for given laser-beam positions $\x_{k,i}\in\partial\Omega$, beam
diameter $\sigma_\x$, time delays $\tau_{k,i}\geq 0$, and angular
departures $\delta_{k,i}=|\theta-\theta_{k,i}|/\sigma_{\theta}$, where
$\theta_{k,i}$ ($0\leq \theta_{k,i} < 2\pi$) and $\sigma_{\theta}$
model the center and the angular spread of the nearly collimated
irradiation from the $(k,i)$-th laser beam. Note that, for both the TS
and MSS configurations a total of
\begin{equation}\label{n_beams}
  N_b = N_qN_s
\end{equation}
laser beams are utilized.

For the TS-type sources we set $N_s=1$ and, generally, $N_q > 1$ (that
is to say, a sequence of $N_q > 1$ single-laser sources is used,
requiring the solution of $N_q$ pairs of forward/adjoint problems per
gradient-descent iteration), while in the proposed MSS-type sources we
let $N_s>1$ and $N_q = 1$ (so that $N_s>1$ laser beams are combined in
a single ``generalized'' source, thus requiring the solution of only
one forward/adjoint problem pair per gradient-descent iteration).  In
each ``sweep'' of the TS method each one of the $N_q>1$ sources is
applied independently of all others, with delays $\tau_{1,i}=0$
($1\leq i\leq N_q$), and readings are recorded at all of the detectors
used~\cite{Prieto2017,Dorn}.  In the proposed MSS method, instead, the
single generalized source ($N_q=1$) is used which incorporates
time-staggered contributions from $N_s>1$ laser beams along
$\partial \Omega$, with time delays $\tau_{k,1}\geq 0$. In view of the
time delays it utilizes, the single forward/adjoint solve used in the
MSS method requires longer computing time than each one of the $N_q>1$
forward/adjoint solves required by the TS method. In all, as
illustrated in Section~\ref{sec:inverseres}, the combined-source
strategy inherent in the MSS approach leads to significant gains in
the overall inversion process without detriment in reconstruction
accuracy.

Both the TS and MSS approaches rely on use of a number $N_d\geq 1$ of
detectors, where the $j$-th detector ($1\leq j\leq N_d$), which is
placed at the point $\x_j \in \partial \Omega$, is characterized by a
measurement operator $G_j=G_j[u](t)$ defined by
\begin{equation}
\begin{split}
\begin{aligned}
  G_j[u]=&\oint_{\partial \Omega}\int_{\hth \cdot \hnu>0}[1-f(\hth
  \cdot \hnu)]
  \hth \cdot \hnu \\
  &\times w\left( \frac{ |\x-\x_j |}{\sigma_d} \right) u(\x,\theta,t)
  d\theta dS
\end{aligned}
\end{split}
\label{eq:OpMed}
\end{equation}
for any given function $u = u(\x,\theta,t)$ defined for
$(\x,\theta,t)\in\Omega\times [0,2\pi)\times [0,T] $. Here, using
equation~\eqref{eq:sourcelabwind} and letting $\sigma_d>0$ denote the
effective area of the detectors, the factor $w( |\x-\x_j |/\sigma_d) $
characterizes the spatial sensitivity of the $j$-th detector and $dS$
denotes the element of area on $\partial \Omega$. Clearly, the
operator $G_j$ quantifies the flux of transmitted photons over the
surface of the detector.  For each generalized source $q_i$ we have a
set of $N_d$ time resolved detector readings. The position and number
of detectors remain fixed throughout the inversion process.

\subsection{Inverse problem}\label{sec:inverse_problem}

The reconstruction of the absorption properties in tissue enables the
identification of tumors~\cite{Zhu2005,Zhu2010,Fujii2016b}, functional
imaging of the brain~\cite{Boas2001,bluestone2001,Arridge1999}, and
characterization of different tissue constituents in medical
imaging. In this work we focus in the reconstruction of the absorption
coefficient only, although the proposed approach can be easily
extended to other reconstruction problems, such as, e.g., the problem
of determining the RTE sources, with application in the related
discipline of fluorescence optical tomography and bioluminescence
tomography~\cite{Klose2009,Ren2010,Klose2005}.  Prior information on
the scattering coefficient $b(\x)$, which can be obtained from high
resolution imaging modalities, is generally assumed for cancer
diagnosis and treatment
monitoring~\cite{Althobaiti2017,Guven2003}. Such prior knowledge
additionally provides (limited) information on the parameter $a(\x)$,
namely, the known absorption coefficient of, say, bone and air, on one
hand, as well as upper and lower bounds on the absorption coefficient
of soft tissue, which can be used to constrain the space of functions
where the minimizer is sought. In view of these considerations, in
what follows we make explicit the dependence of the transport operator
$\mathcal{T}$ and the solution $u$ in eq.~\eqref{eq:Toper} on the
absorption coefficient $a=a(\x)$ by denoting
\begin{equation}\label{eq:Toper_a}
  \mathcal{T}[u] = \mathcal{T}[u,a]= \mathcal{T}[u,a](\x,\theta,t)
\end{equation}
and
\begin{equation}\label{u_of_a}
  u=u[a]=u[a](\x,\theta,t),
\end{equation}
respectively.

We express our inversion problem for the optical parameter $a(\x)$ in
terms of the problem of minimization of the objective function
\begin{equation}
  \Lambda[a]=\sum_{i=1}^{N_q} g_i[u_i],
\label{eq:FObjpr}
\end{equation}
where, for a given absorption coefficient $a$,
\begin{equation}\label{ui_of_a}
  u_i = u_i[a] = u_i[a](\x,\theta,t)
\end{equation}
denotes the solution $u=u_i$ of equation~\eqref{eq:RTE} with $q=q_i$
(increasingly added detail is included in eq.~\eqref{eq:FObjpr} from
left to right concerning the dependence of $u_i$ on $a$ and the
spatial, angular and temporal variables), and where, for a given
number $N_q \times N_d$ of detector measurements $\widetilde{G}_{j,i}$
($N_d$ detector readings $\widetilde{G}_{j,i}$ for each of the $N_q$
generalized sources $q_i$) and using eq.~\eqref{eq:OpMed} $g_i$,
denotes the functional
\begin{equation}
  g_i[u] = \frac{1}{2} \sum_{j=1}^{N_d} \int_0^T (G_j[u]-\widetilde
  {G}_{j,i})^2 dt.
\label{eq:FObj}
\end{equation}

\subsection{Functional derivatives}\label{sec:functional_derivatives}
To minimize the objective function eq.~\eqref{eq:FObjpr} we rely on a
gradient descent algorithm based on use of the functional derivative
$\frac{d\Lambda}{da} [a;\delta a]$ with respect to the absorption
coefficient function $a = a(\x)$ in the direction $\delta a$. Here
$\frac{d}{da}$ denotes Gateaux differentiation~\cite{Hille1974}: for a
given function $a= a(\x)$ and a given perturbation
$\delta a= \delta a(\x)$, the Gateaux derivative of a given functional
$h = h[a]$ in the direction $\delta a$ is defined by
\begin{equation}\label{gateaux}
  \frac{dh}{da}[a; \delta a ] = \lim_{\varepsilon \to 0} \frac{h[a +\varepsilon \delta a]
    - h[a]}{\varepsilon}.
\end{equation}
A similar definition can be given for partial Gateaux derivatives for
an operator $w = w[a](\x,\theta,t)$ (such as, e.g., the
operator~\eqref{eq:Toper}, the solution $u = u[a]=u[a](\x,\theta,t)$
of equation~\eqref{eq:RTE}, etc.):
\begin{equation}\label{gateaux_part}
  \frac{\partial w}{\partial a}[a; \delta a ](\x,\theta,t) = \lim_{\varepsilon \to 0} \frac{w[a +\varepsilon \delta a](\x,\theta,t)
    - w[a](\x,\theta,t)}{\varepsilon}.
\end{equation}

In what follows we utilize Gateaux derivatives of composition of
functionals and operators, for which the chain rule is satisfied. For
example, for the composition $h\circ w [a] = h\big[w[a]\big]$ we have
the chain-rule identity
\begin{equation}\label{eq:chain}
  \frac{d (h\circ w)}{da} \big[a;\delta a\big]= \frac{d h}{
    d w}\left[w[a];\frac{\partial w}{\partial a} \big[a;\delta
    a\big]\right].
\end{equation}
In our context we may illustrate this relationship as follows. As a
variation equal to a real number $\varepsilon$ times a function
$\delta a = \delta a(\x)$ is added to the function $a$, a perturbed
function $(a+\varepsilon\delta a)$ is obtained and, thus, a perturbed
operator value $w[a+\varepsilon\delta a]$. (In our case, the perturbed
operator value could be e.g. the solution $u[a+\varepsilon\delta a]$ of
equation~\eqref{eq:RTE} with absorption coefficient
$(a+\varepsilon\delta a)$; cf. eq.~\eqref{u_of_a}.) In view of the
Gateaux-derivative definition~\eqref{gateaux_part} we obtain
\[
  w[a+\varepsilon\delta a] = w[a]+\varepsilon \frac{\partial
    w}{\partial a}[a; \delta a ] + o(\varepsilon)
\]
where $\frac{o(\varepsilon)}{\varepsilon}\to 0$ as $\varepsilon\to
0$. In other words, the error in the approximation
$w[a+\varepsilon\delta a] \approx w[a]+\varepsilon \frac{\partial
  w}{\partial a}[a; \delta a ]$ is much smaller than $\varepsilon$.
We may thus utilize the approximation
\[
  h\big[w[a+\varepsilon\delta a]\big] \approx h\left[w[a]+\varepsilon \frac{\partial
    w}{\partial a}[a; \delta a ]\right]
\]
in the quotient of increments, of the form~\eqref{gateaux_part}, for
the derivative of the composite function $h\big[w[a]\big]$, which
yields
\[
  \begin{split}
 \lim_{\varepsilon\to 0}\frac{h\big[w[a+\varepsilon\delta a]\big]
     -h\big[w[a]\big]}{\varepsilon} & = \\ \lim_{\varepsilon\to
    0}& \frac{h\big[w[a]+\varepsilon \frac{\partial w}{\partial a}[a;
    \delta a ]\big]-h\big[w[a]\big]}{\varepsilon},
  \end{split}
\]
and, thus, clearly, the right-hand side of~\eqref{eq:chain}, as
desired.

The needed functional derivative of the
objective function~\eqref{eq:FObjpr} is given by
\begin{equation}
  \frac{d \Lambda}{da} = \sum_{i=1}^{N_q} \frac{d (g_i\circ
    u_i)}{da}[a;\delta a].
\label{eq:Gradsumq}
\end{equation}
In order to obtain the derivatives in the sum on the right-hand side
of this equation we apply the chain rule identity~\eqref{eq:chain},
which yields
\begin{equation}
\frac{d (g_i\circ u_i)}{da}[a;\delta a] = \frac{d g_i}{
    d u}\left[u_i[a];\frac{\partial u_i}{\partial a} \big[a;\delta
    a\big]\right],
  \label{eq:AdjointMEthod}
\end{equation}
or, using~\eqref{eq:OpMed} and~\eqref{eq:FObj}, $\frac{d (g_i\circ u_i)}{da}[a;\delta a]=\mathcal{G}[a;\delta a]$ where
\begin{equation}
\begin{split}
\begin{aligned}
  &\mathcal{G}[a;\delta a]\coloneqq \\
  &\int_0^T \oint_{\partial \Omega} \int_{\hth \cdot \hnu>0}
  \sum_{j=1}^{N_d} \left( G_j\big[u_i[a]\big]-
    \widetilde {G}_{j,i} \right) [1-f(\hth \cdot \hnu)] \\
  &\qquad \times \hth \cdot \hnu w\left( \frac{ |\x-\x_j |}{\sigma_d}
  \right) \frac{\partial u_i}{\partial a}[a;\delta a](\x,\theta,t)
  d\theta dS dt.
\end{aligned}
\end{split}
\label{eq:AdjointMEthod2}
\end{equation}

Clearly, in view of eq.~\eqref{eq:AdjointMEthod2}, the
gradients~\eqref{eq:Gradsumq} required by the gradient descent
strategy in a fully discrete context could be produced by evaluating
and substituting in this equation the derivative
$\frac{\partial u_i}{\partial a}[a;\delta a]$, for each (discretized)
absorption coefficient $a$ in the gradient descent process, and for
all (discretized) directions $\delta a$. But, the evaluation of these
partial derivatives, say, by means of a simple finite difference
scheme, requires evaluation of one fully spatio-temporal solution of
the transport eq.~\eqref{eq:RTE} for each direction $\delta a$, which
clearly entails an extremely high, crippling, computational burden. To
avoid this computational expense we rely on the adjoint-method
strategy, which is described in what follows.

\subsection{Fast gradient evaluation via the adjoint method}\label{sec:gradient}
To evaluate the derivative displayed in eq.~\eqref{eq:AdjointMEthod2}
we seek to eliminate the quantity
$\frac{\partial u_i}{\partial a}[a;\delta a]$ from the right-hand side
of this equation. As indicated in what follows, this can be achieved
by considering the initial and boundary problem that is obtained by
differentiation, for the given $a$ and in the direction $\delta a$, of
each one of the three equations in the initial and boundary
problem~\eqref{eq:RTE}. From the first line in~\eqref{eq:RTE}, in
particular, we obtain
\begin{equation}
\begin{split}
\begin{aligned}
  0 = \frac{d\mathcal{T}}{da}\big[u_i[a],a;\delta a
  \big]=&\frac{\partial \mathcal{T}}{\partial u}
  \left[u_i[a],a;\frac{\partial u_i}{\partial a}\big[a;\delta
    a\big]\right] \\&+ \frac{\partial \mathcal{T}}{\partial
    a}\big[u_i[a],a; \delta a \big].
\end{aligned}
\end{split}
\label{eq:RRTEder}
\end{equation}
But, by linearity of $\mathcal{T}$ we have
\begin{equation}
\begin{split}
\begin{aligned}
\frac{\partial \mathcal{T}}{\partial u}\left[u_i[a],a;\frac{\partial u_i}{\partial a}\big[a;\delta a\big]\right]=
\mathcal{T}\left[\frac{\partial u_i}{\partial a}\big[a;\delta a\big],a\right],
\end{aligned}
\end{split}
\label{eq:RRTEdet3}
\end{equation}
and, thus, in view of~\eqref{eq:RTE}, the relation
\begin{equation}
\frac{\partial \mathcal{T}}{\partial a}\big[u_i[a],a;\delta a\big] + 
\mathcal{T}\left[ \frac{\partial u_i}{\partial a}\big[a;\delta a\big],a \right]=0
\label{eq:RRTEdet4}
\end{equation}
results. This relation provides, for each relevant triple
$(\x,\theta,t)$, one linear equation for the two unknowns $u_i[a]$ and
$\frac{\partial u_i}{\partial a}\big[a;\delta a\big]$.

In order to eliminate $\frac{\partial u_i}{\partial a}[a;\delta a]$
from the right-hand side of~\eqref{eq:AdjointMEthod2} we subtract from
from both sides of this identity a ``linear combination with suitable
coefficients'' $\lambda$ of the relation~\eqref{eq:RRTEdet4}---or,
more precisely, an integral of the product of this relation times a
suitable function $\lambda (\x,\theta,t)$ over
$(\x,\theta,t)\in\Omega\times [0,2\pi)\times [0,T]$. (Below we
incorporate additional equations related to the initial and boundary
conditions in~\eqref{eq:RTE} as well.) For notational compactness we
express such integrals in terms of the scalar product notation
\begin{equation}\label{scalar}
  \left\langle v,w \right\rangle= \int_0^T\int_{\Omega}\int_{S^1}
  v(\x,\theta,t)w(\x,\theta,t) d\theta d\x dt
\end{equation}
for any two functions $v$ and $w$ of the variables
$(\x,\theta,t)$. Thus, for a given function $\lambda_i =
\lambda_i[a](\x,\theta,t)$ we obtain from~\eqref{eq:RRTEdet4} the
equation
\begin{equation}
  \left \langle \lambda_i , 
    \frac{\partial \mathcal{T}}{\partial a}[u_i,a; \delta a]
  \right \rangle + \left \langle \lambda_i , 
    \mathcal{T}\left[\frac{\partial u_i}{\partial a}\big[a;\delta a\big],a\right] \right \rangle
  =0,
\label{eq:RRTEdet5}
\end{equation}
which, for a suitably selected function $\lambda_i$ we intend to
subtract from~\eqref{eq:AdjointMEthod2} to achieve the cancellation of
the challenging derivative term.

To select the function $\lambda_i$ that attains such cancellation we
rely on an integration-by-parts calculation to express the second
summand in~\eqref{eq:RRTEdet5} as an integral of a product of two
functions, one of which is precisely
$\frac{\partial u_i}{\partial a}$. Integration by parts of that second
summand leads to a sum of a ``volumetric'' integral $\mathcal{A}$
(namely, an integral over $\Omega\times [0,2\pi)\times [0,T]$) plus a
sum $\mathcal{B} +\mathcal{C}$ of integrals over various portions of
the boundary of this domain:
\begin{equation}\label{eq:int_parts_termabc}
  \left \langle \lambda_i , \mathcal{T}\left[\frac{\partial
        u_i}{\partial a}\big[a;\delta a\big],a\right] \right \rangle = \mathcal{A} +\mathcal{B} +\mathcal{C}
\end{equation}
where
\begin{flalign}\label{eq:A}
&\displaystyle \mathcal{A}[a;\delta a]\coloneqq \int_0^T  \int_{\Omega} \int_{S^1} \frac{\partial u_i}{\partial a}[a;\delta a] \Bigg[-\frac{1}
{c}\frac{\partial \lambda_i}{\partial t} &&
\\\nonumber
&- \hth \cdot \nabla \lambda_i+ (a+b)\lambda_i 
- b\int_{S^1}
\eta(\hth\cdot\hth ') \lambda_i d\theta'\Bigg] d\theta  d\x dt,&&
\end{flalign}

\begin{flalign}
\mathcal{B}[a;\delta a]\coloneqq\int_{\Omega}\int_{S^1} \left[\frac{\partial u_i}{\partial a}[a;\delta a] \lambda_i\right]_0^T d\theta d\x, &&
\label{eq:B}
\end{flalign}
and
\begin{flalign}
  \displaystyle \mathcal{C}[a;\delta a]\coloneqq\int_0^T
  \oint_{\partial \Omega} \int_{S^1} \hth\cdot \hnu \lambda_i
  \frac{\partial u_i}{\partial a}[a;\delta a] d\theta dS dt.&&
\label{eq:C}
\end{flalign}
Subtracting the linear combination~\eqref{eq:RRTEdet5}
from~\eqref{eq:AdjointMEthod2} and using~\eqref{eq:int_parts_termabc}-\eqref{eq:C}
we obtain
\begin{equation}
\begin{split}
\begin{aligned}
\frac{d (g_i\circ u_i)}{da}[a;\delta a] = 
&\mathcal{G} -\mathcal{A} -\mathcal{B} -\mathcal{C}
\\&-\left \langle \lambda_i , 
\frac{\partial \mathcal{T}}{\partial a}[u_i,a;\delta a]
 \right \rangle.
\end{aligned}
\end{split}
\label{eq:AdjointMEthodIII}
\end{equation}
Clearly, the quantity $\frac{\partial u_i}{\partial a}$
in~\eqref{eq:AdjointMEthodIII} will be eliminated, as desired, if and
only if
\begin{equation}
  \mathcal{A} +\mathcal{B} +\mathcal{C} = \mathcal{G},
\label{eq:AdjointMEthod4}
\end{equation}
since the last term on the right-hand side
of~\eqref{eq:AdjointMEthodIII} does no contain
$\frac{\partial u_i}{\partial a}$. Once we select $\lambda_i$ such
that~\eqref{eq:AdjointMEthod4} is satisfied, and using the Gateaux
derivative relation
\begin{equation}
\begin{split}
\begin{aligned}
 \frac{\partial \mathcal{T}}{\partial a}[u_i,a;\delta a]=\delta a(\x) u_i[a](\x,\theta,t),
\end{aligned}
\end{split}
\label{eq:kdeltalgat}
\end{equation}
the  expression
\begin{equation}
\displaystyle \frac{d (g_i\circ u_i)}{da}[a;\delta a] =
 -\langle \lambda_i[a] , \delta a  u_i[a] \rangle 
\label{eq:FDwin2}
\end{equation}
for the functional derivative, which does not contain the challenging
term $\frac{\partial u_i}{\partial a}$, results
from~\eqref{eq:AdjointMEthodIII}.

To obtain the solution $\lambda_i=\lambda_i[a](\x,\theta,t)$ of
eq.~\eqref{eq:AdjointMEthod4} we note that, in view of the spatial
integration domains in eqs.~\eqref{eq:AdjointMEthod2}
and~\eqref{eq:A}-\eqref{eq:C}, eq.~\eqref{eq:AdjointMEthod4} is
satisfied if and only if (i)~$\mathcal{A}=0$, (ii)~$\mathcal{B}=0$ and
(iii)~$\mathcal{C}-\mathcal{G}=0$. Equation (i) is satisfied provided
the term in brackets in the integrand of~\eqref{eq:A}, which will be
denoted by $\mathcal{T}^*\big[\lambda_i[a],a\big]$ in what follows,
equals zero. Relation (ii) is satisfied by imposing the appropriate
``final''condition $\lambda_i(\x,\theta,t=T)=0$, since, in view
of~\eqref{eq:RTE}, we have $\frac{\partial u_i}{\partial a} = 0$ for
$t=0$.  In order to fulfill point~(iii), finally, we decompose the
integral~\eqref{eq:C} into two integrals, $\mathcal{C}_+$ and
$\mathcal{C}_-$, where integration ranges in the $\theta$ variable are
restricted to angular domains $\hth\cdot \hnu >0$ and
$\hth\cdot \hnu <0$, respectively. The integrand in the difference
$\mathcal{C}_+ -\mathcal{G}$, which is only integrated over the
angular domain $\hth\cdot \hnu >0$, equals the product of the common
factor $F = \frac{\partial u_i}{\partial a}\hth\cdot \hnu$ and the
difference
$P=\lambda_i-\sum_{j=1}^{N_d} \Big( G_j[u_i]-\widetilde{G}_{j,i} \Big)
\times [1-f(\hth \cdot \hnu)] w\left( \frac{ |\x-\x_j |}{\sigma_d}
\right)$. To incorporate the summand $\mathcal{C}_-$ under the same
$\hth\cdot \hnu >0$ integration range, in turn, we first utilize the
Fresnel boundary condition
$\frac{\partial u_i}{\partial a}(\x,\theta,t)=f(\hth \cdot \hnu)
\frac{\partial u_i}{\partial a}(\x,\theta_r,t)$
($(\x,\theta) \in \Gamma_-$) that results from differentiation of the
boundary condition in eq.~\eqref{eq:RTE}, and we thus obtain
\begin{flalign*}
 \mathcal{C}_-[a;\delta a]\coloneqq&\int_0^T \oint_{\partial \Omega}
\int_{\hth\cdot\hnu<0} \hth \cdot \hnu \lambda_i(\x,\theta,t) f(\hth\cdot\hnu)\\&\times \frac{\partial
  u_i}{\partial a}[a;\delta a](\x,\theta_r,t) d\theta dS dt.&&
\end{flalign*}  
Then,
incorporating the change of variables
$\theta=2\theta_{\nu}-\theta_r+\pi$, so that
$\hth\cdot\hnu = \cos(\theta - \theta_\nu) = -\cos(\theta_r -
\theta_\nu) = -\hth_r\cdot \hnu$  we obtain
\begin{flalign*}
 \mathcal{C}_-[a;\delta a]\coloneqq& - \int_0^T \oint_{\partial
  \Omega} \int_{\hth_r\cdot \hnu>0} \hth_r\cdot \hnu \lambda_i(\x,2\theta_{\nu}-\theta_r+\pi,t) 
\\&\times f(\hth_r\cdot\hnu) \frac{\partial u_i}{\partial a}[a;\delta
a](\x,\theta_r,t) d\theta_r dS dt.
&&
\end{flalign*}  
Substituting the dummy variable $\theta_r$ by $\theta$ in this
equation, the angular argument in $\lambda_i$ becomes
$2\theta_{\nu}-\theta+\pi$ which coincides with $\theta_r$, and, thus,
calling $Q(\x,\theta_r,t)=f(\hth \cdot \hnu) \lambda_i(\x,\theta_r,t)$
we obtain
\begin{flalign*}
 \mathcal{C}_-[a;\delta a]\coloneqq& - \int_0^T \oint_{\partial
   \Omega} \int_{\hth\cdot \hnu>0} \hth \cdot \hnu Q(\x,\theta_r,t)\\ &
 \times \frac{\partial u_i}{\partial a}[a;\delta
a](\x,\theta,t) d\theta dS dt.
&&
\end{flalign*}
Combining this result with the term $\mathcal{C}_+-\mathcal{G}$ we
obtain
\begin{flalign*}
 (\mathcal{C}-\mathcal{G})[a;\delta a]\coloneqq& \int_0^T \oint_{\partial
   \Omega} \int_{\hth\cdot \hnu>0} F\times [ P(\x,\theta,t)\\&
 -Q(\x,\theta_r,t)]
 d\theta dS dt,
&&
\end{flalign*}
and, thus, (iii)~is satisfied provided $P-Q=0$.

In summary, denoting
$\mathcal{T}^*\big[\lambda_i[a],a\big]=-\frac{1} {c}\frac{\partial
  \lambda_i} {\partial t}- \hth \cdot \nabla \lambda_i + (a+b)\lambda_i-
b\int_{S^1}\eta(\hth\cdot\hth ') \lambda_i d\theta'$, we have shown
that the conditions~(i), (ii) and~(iii) are satisfied provided the
corresponding ``adjoint'' problem
\begin{equation}
\begin{split}
\begin{aligned}
  &\mathcal{T}^*\big[\lambda_i[a],a\big]=0, \; (\x,\theta)
  \in \Omega\times [0,2\pi)\\
  &\lambda_i(\x,\theta,t=T)=0,\; (\x,\theta)
  \in \Omega\times [0,2\pi),\quad\mbox{and}\\
  &\lambda_i(\x,\theta,t) = f(\hth \cdot \hnu)
  \lambda_i(\x,\theta_r,t)+\sum_{j=1}^{N_d} \Big( G_j[u_i] \\&\;\;
  -\widetilde {G}_{j,i} \Big) \times [1-f(\hth \cdot \hnu)] w\left( \frac{
      |\x-\x_j |}{\sigma_d} \right), (\x,\theta) \in \Gamma_+
\end{aligned}
\end{split}
\label{eq:AdjointProblem}
\end{equation}
hold. Thus, the function $\lambda_i(\x,\theta,t)$ needed in
eq.~\eqref{eq:FDwin2} can be obtained by solving the \textit{adjoint
  back transport} problem~\eqref{eq:AdjointProblem} in the time
interval $T \geq t \geq 0$ with homogeneous final data at time
$t=T>0$.  Once the function $\lambda_i$ has been obtained the
component of the functional gradient~\eqref{eq:FDwin2} in the
direction $\delta a$ can inexpensively be obtained by integration,
which, in view of~\eqref{scalar}, may be expressed in the form
\begin{equation}\label{gradient_fin}
  \frac{d (g \circ u)}{da}[a;\delta a] =-\int_0^T \int_{\Omega} \int_{S^1} \lambda u \delta
  a d \theta d\x dt.
\end{equation}

The correctness and accuracy of the proposed approach for gradient
evaluation are demonstrated in the following section via comparisons
with direct finite-difference gradient computations.

\subsection{Verification and accuracy assessment of the functional-derivative expression~\eqref{eq:FDwin2}}
\label{sec:gradc}
In this section we present numerical verifications and accuracy
assessments for the functional derivative
expression~\eqref{gradient_fin}, with $\lambda_i$ given as the
solution of the adjoint problem~\eqref{eq:AdjointProblem}. To do this
we consider a problem of the type~\eqref{eq:RTE} with Fresnel boundary
conditions, described in what follows---so as to illustrate, in
particular, the ability of the functional-derivative expression to
produce correct gradients in this case, for which corresponding
adjoint treatments were not previously available. As a basis for
comparison we obtain numerical gradients produced by direct use of the
finite-difference approximation
\begin{equation}
  \frac{d (g \circ u)}{da}[a;\delta a]^{FD} \sim
  \frac{g\big[u[a+\varepsilon \delta a]\big]
    -g\big[u[a]\big]}{\varepsilon}
\label{eq:ObjFD}
\end{equation}
for a given direction $\delta a(\x)$ and a suitable small value of
$\varepsilon$.  The index of refraction $n_{\Omega}=1.4$ is assumed in
the spatial domain
$\Omega=[x_{\text{min}},x_{\text{max}}]\times[y_{\text{min}},y_{\text{max}}]=[0,3]\times[0,3]$
and with $n_0=1$ outside $\Omega$.  For simplicity we use
$\delta a=1$, with given spatially constant values $a(\x)=a$ and
$b(\x)=b$ of the absorption and scattering coefficients, and, without
loss of generality, we consider a single generalized source $q_1 = q$
for both a TS source case (a single laser incident beam located at
$\x_s=(1.5,0.0)$) and an MSS source case (assumed to consist of the
combination of four laser incident beams, one located at the center of
each one of the sides of the square domain $\Omega$). (Further details
on the modeling of sources can be found at the end of
sec.~\ref{sec:forwsim}, and eq.~\eqref{eq:sourcelab}.)  For these
tests we employ a single detector placed at $\x_d=(0.0,0.75)$. The
time delays required by the MSS method are selected for these examples
by enforcing a $50$ps time-shift between successive laser start
times. A total duration of $60$ps was used for each single pulse, and
the system was evolved for both TS and MSS cases up to a final time of
$600$ps.  A mesh with $N_x=N_y=200$, $M=32$ discrete directions and
$T=60000$ time steps was used for the solution of the RTE and its
adjoint.  The relative error
\begin{equation}
 e=\displaystyle \frac{\left|\frac{d (g\circ u)}{da}[a;\delta a]^\mathrm{Adj}- \frac{d (g\circ u)}{da}[a;\delta a]^\mathrm{FD}\right|}{|\frac{d (g\circ u)}{da}[a;\delta a]^\mathrm{Adj}|},
\label{eq:Errgrad}
\end{equation}
was used to quantify the quality provided by the proposed adjoint
gradient expression, where
$\frac{d (g\circ u)}{da}[a;\delta a]^\mathrm{Adj}$ and
$\frac{d (g\circ u)}{da}[a;\delta a]^\mathrm{FD}$ denote the adjoint
and finite difference derivatives, respectively; the value
$\varepsilon=0.0001$ was used to produce the finite difference
approximation~\eqref{eq:ObjFD}.

\begin{table}[h!]
\caption{Functional derivative differences}
\vspace{-0.3cm}
\begin{center}
\begin{tabular}{cccccc}
\hline
$a[1/cm]$ & ~ & $b[1/cm]$ ~ & $g$ ~ & $e_{\text{TS}}$  ~ & $e_{\text{MSS}}$ \\
\hline
$0.35$ & ~ & $80$ ~ &$0.9$ ~ & $0.00008$  ~ & $0.00009$ \\
$0.35$ & ~ & $20$ ~ &$0.0$ ~ & $0.00008$  ~ & $0.00097$ \\
$0.35$ & ~ & $8.0$ ~ &$0.0$ ~ & $0.00027$  ~ & $0.00016$ \\
$0.35$ & ~ & $0.1$ ~ &$0.0$ ~ & $0.00009$  ~ & $0.00026$ \\
\hline
\end{tabular}
\label{tab:grads}
\end{center}
\end{table}
Table~\eqref{tab:grads} demonstrates the agreement observed between
the values of the functional derivative produced by the
finite-difference and adjoint methods under various transport regimes,
including several values of the scattering coefficient $b$ and
anisotropy coefficient $g$, and under both the TS and the MSS
configurations; errors of similar magnitudes were obtained for a wide
range of values of the parameters $a$, $b$ and $g$. The excellent
agreement observed in all cases suggests that the very large
improvements in computational speed provided by the adjoint method,
which would amount to a factor of the order of $(N_x+1) \times (N_y+1) \simeq 40,000$ for
the evaluation of the full gradient in the present example, do not
impact upon the accuracy in the gradient determination.

\subsection{Numerical functional gradient calculation}
\label{sec:gradc}

All of the numerical gradients utilized in this paper were obtained by
solving forward and the adjoint problems, followed by use of a
discrete version of eq.~\eqref{gradient_fin} for a number of
perturbation functions $\delta a$---each one of which is selected to
provide a variation of the absorption coefficient at and around one of
the discretized spatial coordinate points
$\x_{\ell_1,\ell_2} \in \Omega$ in the discretization
$\x_{\ell_1,\ell_2} =(x_{\text{min}}+ [\ell_1-1]\Delta x)\hat x
+(y_{\text{min}}+[\ell_2-1]\Delta y)\hat y$, $\ell_1=1,\ldots,N_x+1$,
$\ell_2=1,\ldots,N_y+1$ of the domain $\Omega$.  The perturbation
$\delta a$ is selected as a pyramid-shaped function which equals one
at a point $\x_{\ell_1,\ell_2} \in \Omega$, and becomes zero at and
beyond the first neighbors in the discrete grid. At the discrete
level, each pyramid-shaped function is approximated by a product of
Kronecker delta functions
$\delta a_{\ell_1,\ell_2}=\delta_{r,\ell_1}\times\delta_{s,\ell_2}$.
Thus, denoting by $\nabla_a g(\x_{\ell_1,\ell_2})$ the value of the
functional gradient in the direction $\delta a_{\ell_1,\ell_2}$, the
discrete version of eq.~\eqref{gradient_fin} is given by
\begin{equation}
\nabla_a g(\x_{\ell_1,\ell_2})  \sim -\sum_{m,j}  \lambda_{\ell_1,\ell_2,m,j}u_{\ell_1,\ell_2,m,j} \Delta \theta \Delta x \Delta y \Delta t,
\label{eq:FDwin22}
\end{equation}
where
$\lambda_{\ell_1,\ell_2,m,j}\sim
\lambda(\x_{\ell_1,\ell_2},\theta_m,t_j)$ and where
$u_{\ell_1,\ell_2,m,j} \sim u(\x_{\ell_1,\ell_2},\theta_m,t_j)$.  Note
that eq.~\eqref{eq:FDwin22} represents the functional derivative for a
single direction $\delta a_{\ell_1,\ell_2}$, corresponding to the
$(\ell_1,\ell_2)$ component of the functional gradient. As a result of
the adjoint method, the evaluation of eq.~\eqref{eq:FDwin22} for all
$(\ell_1,\ell_2)$ requires only one forward and one adjoint transport
simulation for each generalized source $q=q_i$---a calculation which,
if performed by direct use of eq.~\eqref{eq:ObjFD} requires a much
larger number $(N_x+1)\times (N_y+1)$ of forward transport simulations
and associated overwhelming computational cost.  It must be noted,
however, that the adjoint method requires storage in memory of full
forward and adjoint transport solutions. The solver algorithm proposed
in Section~\ref{sec:nmethods} is well suited for parallel distributed
systems, as it simultaneously provides computational speed and
distributed memory availability.

\begin{figure}[h!]
\centering
  \includegraphics[width=0.75\linewidth]{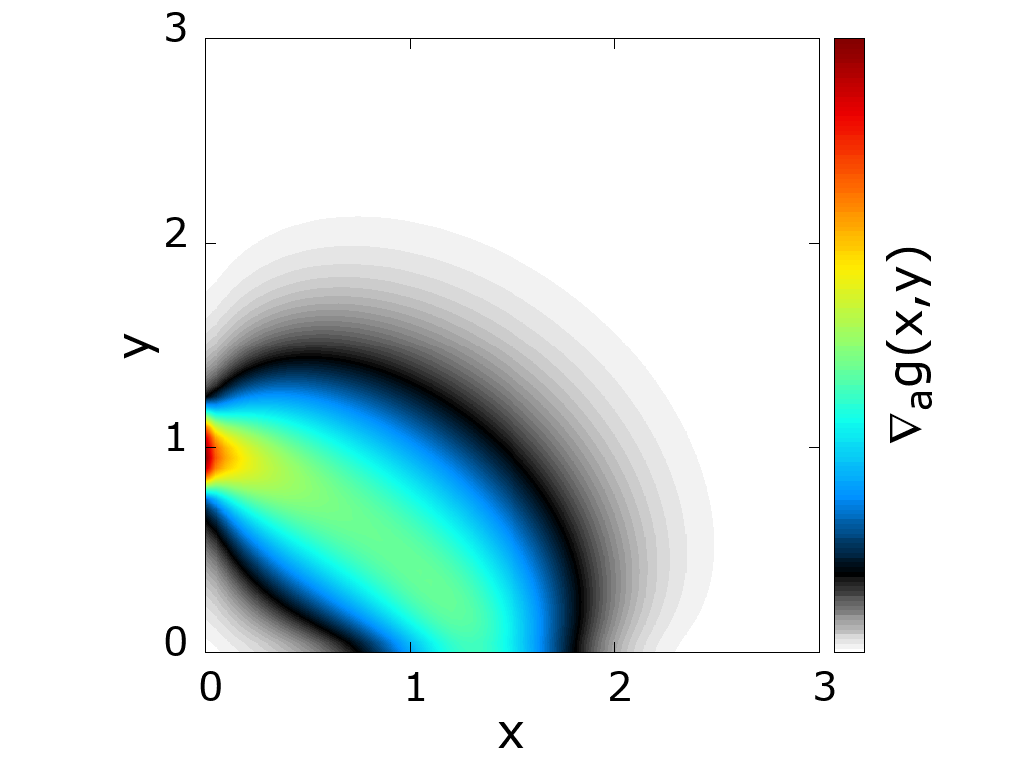}
  \caption{(Color online).  Full spatial gradient
    eq.~\eqref{eq:FDwin22} (in arbitrary units) for a single
    source-detector pair, located at $\x_s=(1.5,0)$ and $\x_d=(0,1.0)$
    respectively, for selected data values $\widetilde {G}_{1,1}$ (one
    source and one detector).}
 \label{fig:gradient}
\end{figure}

As an illustration, Figure~\ref{fig:gradient} displays the full spatial
gradient $\nabla_a g(\x_{\ell_1,\ell_2})$ for
$1\leq \ell_1\leq (N_x+1)$ and $1\leq \ell_2\leq (N_y+1)$, for certain
assumed values $\widetilde G_{j,i}$, with a single source and a single
detector placed at $\x_s=(1.5,0)$ and $\x_d=(1.0,0)$ respectively.

\section{Parallel FC--DOM numerical implementation for the Radiative
  Transfer Equation}
\label{sec:nmethods}

The numerical treatment of the time dependent RTE requires a
discretization of all variables in phase space, namely, the spatial,
directional and temporal variables. In this section we present a
parallel algorithm for the numerical solution of the RTE on the basis
of such a discrete grid in phase space. As evidenced by the approach
used, and demonstrated by means of numerical experiments in
Sections~\ref{decomp_parallel} and~\ref{sec:forwsim}, the proposed
algorithm enjoys high order accuracy for smooth solutions as well as
high parallel efficiency.

\subsection{Velocity-domain discretization}

We discretize the RTE with respect to the velocity variable
$\mathbf{v}=c\hth$ by means of the discrete ordinates method.  To do
this a set of $M$ discrete directions
$\hth(\theta_m)=\hth_m=(\xi_m,\eta_m)$ ($1\leq m\leq M$) is utilized
where the direction cosines are given by
$\xi_m=\hat{x} \cdot \hth_m=\cos({\theta_m})$ and
$\eta_m=\hat{y} \cdot \hth_m=\sin({\theta_m})$ in terms of the
Cartesian unit vectors $\hat{x}$ and $\hat{y}$, where
\begin{equation}
\theta_m=\frac{2\pi(m-1)}{M}, \; \; m=1,2,\ldots,M.
\label{eq:RTETrapz}
\end{equation}
The necessary angular integrations are produced by means of the
trapezoidal rule using the associated quadrature weights
$w_{m}=2\pi/M$.  Given that the specific intensity $u$ is
$2\pi$-periodic in the angular variable $\theta$, and provided the
transport solution is sufficiently smooth, the use of the trapezoidal
rule gives spectral accuracy for integration with respect to this
variable, as illustrated in Figure~\ref{fig:convman}.

Letting $u_m=u(\x,\theta_m,t)$, the semidiscrete version of the
differential equation in~\eqref{eq:RTE} translates into the system of
equations
\begin{equation*}
\begin{aligned}
\begin{split}
\frac{1}{c}  \frac{\partial  u_m}{\partial t}  
 + \hth_m
\cdot  \nabla u_m + (a+b) u_m&   \\
-b\sum_{m'=1}^M w_{m'}
p_{m,m'} \, u_{m'} &=0, \; \; m=1,2,\ldots,M.
\end{split}
\end{aligned}
\label{eq:RTEdom}
\end{equation*}
To ensure energy conservation, and thus enable the stability of the
numerical scheme, the phase function eq.~\eqref{eq:Henyey-Greeinstein}
is normalized in such a way that its {\em numerical} integral, as
produced, in our case, by means of the trapezoidal rule, equals
one~\cite{Kim1988,Liu2002}.

\subsection{Time propagation}
\label{subsec:ABtp}
To evolve the RTE solution in time with high order accuracy, we employ
the fourth order Adams--Bashforth time-stepping
method~\cite{Bulirsch_1991}, which, in the present context, offers a
reasonable compromise between accuracy and stability.  Calling
\begin{equation}\label{t-step}
  u_m^k=u_m^k(x,y)=u(\x,\theta_m,t^k),\quad  t^{k}=(k-1)\Delta t,
\end{equation}
the time dependent RTE equation is evolved from the initial condition
via the relations
\begin{equation}
u_m^{k+1} =  u_m^k + c\Delta t \sum_{\ell=0}^3 \chi_{\ell} \mathcal{L} ( u^{k-\ell}_m)  
\label{eq:RTEAB4}
\end{equation}
\noindent where
\begin{equation}
  \mathcal{L} ( u_m^k) = b\sum_{m'=1}^{M} w_{m'} p_{m,m'} u_{m'}^k-(b+a)u_m^k-\hth_m \cdot \nabla u_m^k \notag,
\label{eq:RTEAB4L}
\end{equation}
with fourth order Adams--Bashforth coefficients given by
$\chi_0=55/44$, $\chi_1=-59/24$, $\chi_2=37/24$ and
$\chi_4=-3/8$. Throughout this work the four initial time-steps were
set to vanish, as befits incident fields that smoothly ramp-up from
zero. This arrangement corresponds to the actual experimental setup in
time-dependent optical tomography. But we note that arbitrary initial
conditions can be treated within our context, by proceeding e.g. as
recommended in Section 5 of ref.~\cite{Bruno2016} in a related
context.

\subsection{Discretization over the spatial domain: the FC--DOM method}
\label{subsec:FCGram}

For simplicity, in this work we consider a square spatial domain
\begin{equation}\label{square}
  \Omega=\{ \x=(x,y) \in [x_{\text{min}},x_{\text{max}}]\times
  [y_{\text{min}},y_{\text{max}}] \},
\end{equation}
although general curvilinear domains can be treated similarly;
cf. e.g.~\cite{Bruno2016,Amlani2016,Bruno2019}.  We discretize the
spatial domain by means of a uniform grid with grid-sizes $\Delta x$
and $\Delta y$ along the $x$ and $y$ directions, respectively;
throughout this paper we have used $\Delta x = \Delta y$. We denote by
$\mathcal{D}$ denote the discretized version of the spatial domain
$\Omega$: $\mathcal{D} = \{ (x_i,y_j) \ |\ i=1,\ldots,N_x+1$ and
$j=1,\ldots,N_y+1\}$. In what follows we describe the method used for
evaluation of spatial derivatives in the $x$ direction, for which we
use the grid
\begin{equation*}
\begin{aligned}
  &x_i =x_{\mathrm{min}}+(i-1)\Delta x, \quad\mbox{with}\\
  &\Delta x =\frac{x_{\text{max}}-x_{\text{min}}}{N_x} = \frac{x_{N_x+1} -
    x_1}{N_x};
\end{aligned}
\end{equation*}
the $y$ derivatives are, of course, handled similarly.

We produce the necessary spatial derivatives by means of the Fourier
Continuation method (FC)~\cite{Bruno2010}, that provides
low-dispersion approximations with high-order accuracy on the basis of
Fourier expansions for general (non periodic) functions.  For a
function defined on an interval in the real line, the FC method
utilizes a smooth and periodic extension of the given function into an
extended interval, for which a regular Fourier series is then
obtained, which accurately approximates the extended periodic function,
and, thus, in particular, the given function in the original interval.
The FC method has extensively been studied and utilized; see
e.g.~\cite{Albin2011,Amlani2016,Bruno2019,Gaggioli2019,Fontana2020} 
and references therein.  In what follows we briefly review this method 
in the context of the RTE considered in the present contribution.

\begin{figure}[h!]
\centering
  \includegraphics[width=0.8\linewidth]{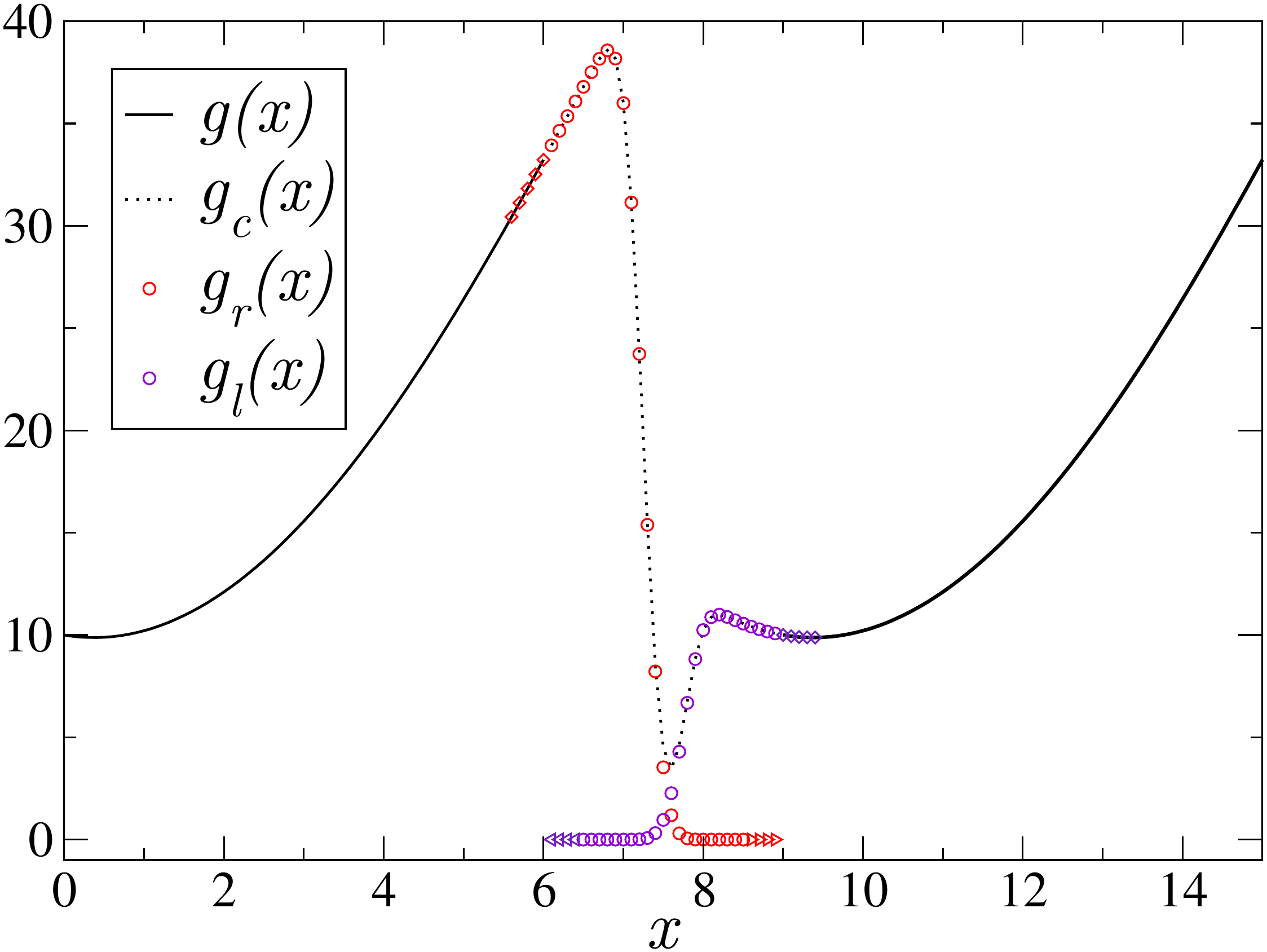}
  \caption{Fourier Continuation of the function
    $g(x)=12+x^2 - e^{x/3}$. For this example, the original function
    is defined for $x\in[0,6]$, and a copy of it, displaced to the
    interval $[9,15]$, is also included. The continuations to zero,
    $g_l(x)$ and $g_r(x)$, are shown in circles. The dotted line is
    the periodic continuation obtained. The diamonds are the matching
    points, with $d_l=d_r=5$, and the triangles are the $E=4$ extra
    points used in this example.}
 \label{fig:Continuedg}
\end{figure}

In order to produce the Fourier Continuation of a given function, such
as e.g. $g(x)=u^k_m(x,y_j)$ for a given $y_j$ in the
$y$-discretization mesh, we consider the discrete vector
$\mathbf{g}=[g_1,\ldots,g(x_i),\ldots,g_{N_x+1}]^T$ of values of the
function $g$. Utilizing a few of the leftmost and rightmost entries of
$\mathbf{g}$, called the ``matching values'' in what follows, the
algorithm produces a continuation vector $\mathbf{g}_c$ that
corresponds to the discrete function values of the desired
continuation function $g_c$---which, once the vector $\mathbf{g}_c$ is
available, can be obtained by an application of the Fast Fourier Transform 
(FFT) algorithm.  To obtain the vector $\mathbf{g}_c$ the algorithm at first smoothly
extends the aforementioned matching values to zero, towards the left
and right, respectively, as illustrated in
Figure~\ref{fig:Continuedg}. To achieve the desired extensions to
zero, certain ``projections'' are used (which project the first $d_l$
function values $(g_1,g_2,\ldots,g_{d_l})$ and the last $d_r$ function
values $(g_{N_x+2-d_r},\ldots,$ $g_{N_x+1})$ onto polynomial bases), and
then, precomputed continuations to zero are utilized for each
polynomial in the bases. Here $d_l$ and $d_r$ denote small integers;
throughout this paper we have used $d_l = d_r =5$.

In detail, the continuation procedure can be summarized in the
following four steps~\cite{Amlani2016}:
\begin{enumerate}
\item The $d_l$ and $d_r$ matching values $g_1,g_2,\ldots,g_{d_l}$ and
  $g_{N_x+2-d_r},g_{N-d_r},\ldots,g_{N_x+1}$ are projected on Gram
  polynomial bases.
\item Smooth continuations to zero $\mathbf{\widetilde{g}}_l$ and
  $\mathbf{\widetilde{g}}_r$ are produced on the basis of accurately
  precomputed extensions to zero~\cite{Albin2011,Amlani2016} of each one of
  the polynomials in the Gram basis.
\item Two new vectors $\mathbf{g}_l$ and $\mathbf{g}_r$ are generated
  which expand the dimension of $\mathbf{\widetilde{g}}_l$ and
  $\mathbf{\widetilde{g}}_r$ to $C+E$ by merely adding zero entries to the
  vector, where $E$ is a number of extra zeroes used. This expansion
  is used to obtain vectors whose dimension can be factored into small
  primes, leading to efficient FFT evaluation.
\item The discrete continued vector $\mathbf{g}_c$ containing
  $N_p=N_x+1+C+E$ components is given by
  \begin{equation*}
\mathbf{g}_c=
\begin{bmatrix}
    \mathbf{g} \\
    \mathbf{g}_l+\mathbf{g}_r  \\
\end{bmatrix}
,
\label{eq:FCStepsCondensed}
\end{equation*}
and the continued function $g_c$ is obtained from $\mathbf{g}_c$ via the FFT.
\end{enumerate}

Figure~\ref{fig:Continuedg} illustrates the FC method and the resulting
continuation function $g_c$ for the non periodic function
$g(x)=12+x^2 - e^{x/3}$ in the interval $[0,6]$ using $N=60$ points in
the grid, $C=25$ continuation points (dotted line), $d_l=d_r=5$
matching points (shown with diamonds) and $E=4$ extra points. The
continuation vector $\mathbf{g}_c$ contains $N_p = N_x + 1 + C + E=90$
entries, which can be factored into small prime numbers,
$N_p=2\times 3 \times 3 \times 5$. As suggested above, the number $E$
of extra points is selected so as to ensure that $N_p$ equals the
product of small prime factors; in practice we enforce that the prime
factors in the factorization $N_p=\prod_{j=1}^{F} p_{j}$ satisfy
$p_1\leq p_2\leq \ldots \leq p_{F} \leq 5$---leading to a efficient
evaluation of Fourier coefficients via the FFT.

Using the Fourier coefficients obtained by means of the FFT, as
indicated above, the continuation $g_c$ is given by
\begin{equation}
g_c(x)=\sum_{k={-N_p/2}}^{N_p/2}a_k \,  \exp \left( 2\pi i k\frac{(x-x_{\text{min}})}{b}\right)
\label{eq:RTEFCSum}
\end{equation}
(where $i$ denotes the imaginary unit, and where $b=[N_p-1]\Delta x$
denotes the period of the continued function).  We additionally employ
an exponential filter which, without deterioration in the
accuracy~\cite{Albin2011}, ensures the stability and robustness of the
method. The filtered coefficients are given by
\begin{equation}
\hat{a}_k=\exp \left(-\alpha \left|\frac{2k}{N_p}\right|^{2\beta}\right)a_k \, 
\label{eq:FFilter}
\end{equation}
for adequately selected values of $\alpha$ and $\beta$;
following~\cite{Albin2011} throughout this paper we use the values
$\alpha=2$ and $\beta=55$ in conjunction with the fourth order
Adams--Bashforth method.  The necessary derivatives of the function
$g$ are obtained by direct differentiation of the filtered version of
the continuation function (which, for notational simplicity, will also
be called $g_c$):
\begin{equation} \displaystyle \frac{d g(x)}{dx} \sim \frac{d
    g_c(x)}{dx}=\sum_{k={-N_p/2}}^{N_p/2} \frac{2\pi i k}{b} \hat{a}_k
  \, e^{2\pi i k\frac{(x-x_{\text{min}})}{b} }.
\label{eq:RTEFCSumder}
\end{equation}

\subsection{Domain decomposition and parallel
  implementation\label{decomp_parallel}}

\begin{figure}[h!]
  \centering
  \includegraphics[width=\linewidth]{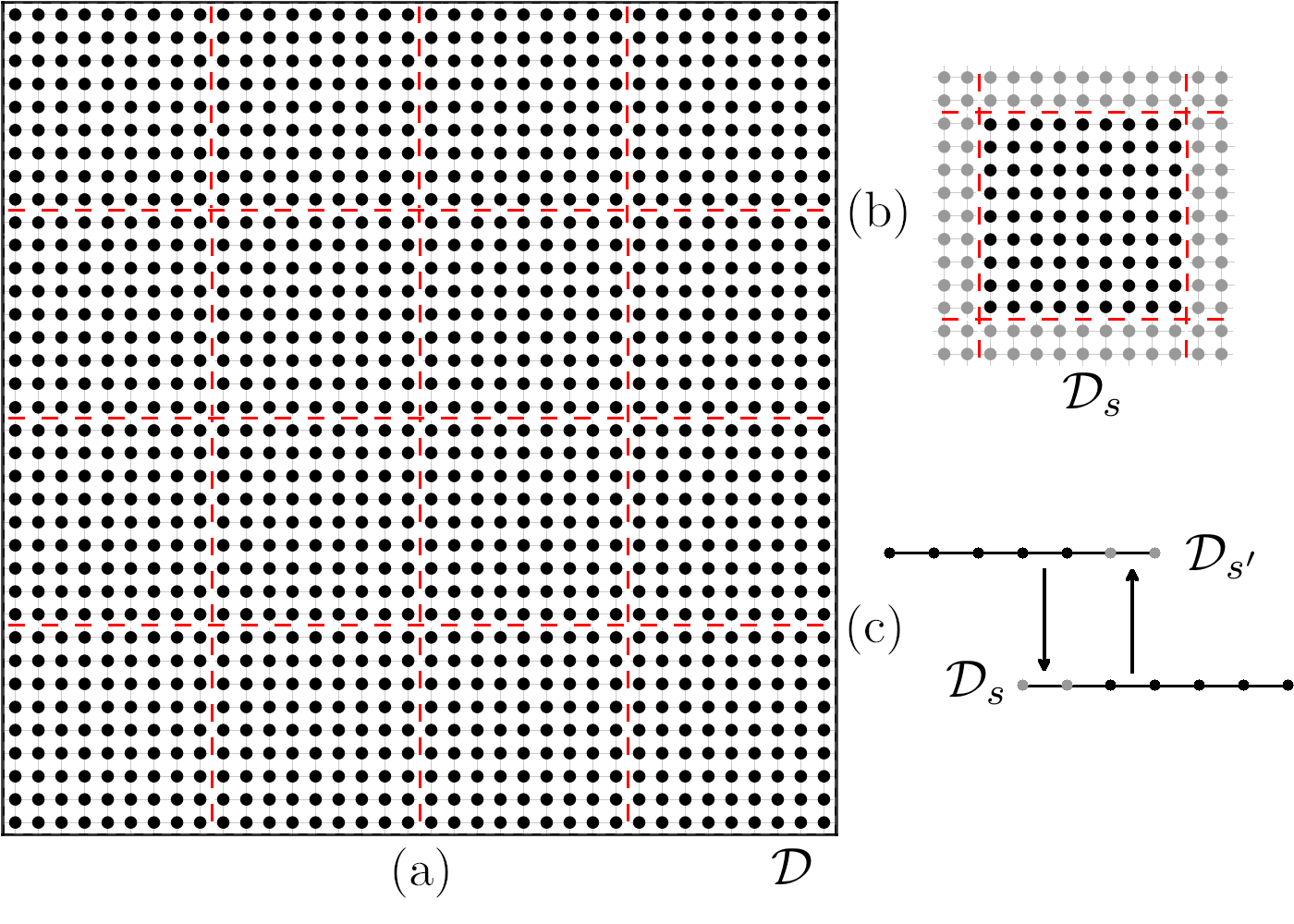}
  \caption{(Color online). (a) The domain $\mathcal{D}$ is decomposed
    as a union non-overlapping subdomains
    $\widetilde{\mathcal{D}}_s$. (b)~Each subdomain
    $\widetilde{\mathcal{D}}_s$ extended by two ``fringe points'' in
    the $x$ and $y$ directions (gray circles), resulting in the
    overlapping subdomains $\mathcal{D}_s$.  (Subdomains adjacent to
    the physical boundary $\Gamma$ are only extended toward the
    interior of $\mathcal{D}$.)  (c) The fringe points are used to
    enable exchange of information between neighboring subdomains
    $\mathcal{D}_s$ and $\mathcal{D}_{s'}$: in the one-dimensional
    cartoon, process $s$ sends to its neighbor $s'$ the $u$ value at
    the third and fourth points and it receives from $s'$ the $u$
    value at the first two points on the $s$ grid.}
 \label{fig:bechange}
\end{figure}

In order to solve the RTE on parallel systems we decompose the
discrete domain $\mathcal{D}$ introduced in
Section~\eqref{subsec:FCGram} as a union
$\mathcal{D} = \cup_{s=1}^{N_{c}}\mathcal{D}_s$ of a number $N_{c}$ of
overlapping subdomains $\mathcal{D}_s$ ($1\leq s\leq N_c$), with
corresponding interior (non-overlapping) regions
$\widetilde{\mathcal{D}}_s$, as illustrated in panels~(a), (b) and (c)
in Figure~\ref{fig:bechange}. The overlapping domains $\mathcal{D}_s$
equal the union of corresponding sets $\widetilde{\mathcal{D}}_s$
(shown as sets of black points partitioned along red dashed lines in
panel~(a)), and a set of ``fringe points'' (shown in gray in
panels~(b) and~(c)).

Each one of the subdomains $\mathcal{D}_s$, which contains $N_{x,s}$
and $N_{y,s}$ points along the $x$ and $y$ directions,
respectively, is assigned to one processing core---so that, in
particular, $N_c$ equals the number of processing cores used. Note
from Figure~\ref{fig:bechange}~(b) (and the one-dimensional cartoon
Figure~\ref{fig:bechange}~(c)) that neighboring subdomains
$\mathcal{D}_s$ overlap by four grid points along the $x$ and
$y$ directions.  In spite of the depiction in
Figure~\ref{fig:bechange}, which only includes square subdomains, in
general, rectangular subdomains need to be used for certain values of
$N_c$.

The subdomain
$\mathcal{D}_s = [x_{\text{min},s},x_{\text{max},s}]\times
[y_{\text{min},s},y_{\text{max},s}]$ contains grid points of $x$- and
$y$-coordinates $x_i=x_{\min,s}+(i-1)\Delta x$ and
$y_j=y_s(j)=y_{\min,s}+(j-1)\Delta y$, respectively.  The algorithm
proceeds by time stepping the quantities $u_m^k(x_i,y_j)$
(cf.~\eqref{t-step}) in parallel on each subdomain $\mathcal{D}_s$. At
the end of each time step, a boundary exchange of four points is
performed between neighboring subdomains as indicated in the caption
of Figure~\ref{fig:bechange}. The proposed algorithm for the evolution
of the time dependent RTE in parallel is summarized in
Algorithm~\ref{algfc}.

\begin{algorithm}
\caption{FC--DOM Parallel Algorithm}\label{algfc}
\begin{algorithmic}[1]
\State  Generate the domain decomposition.
\State Assign one subdomain $\mathcal{D}_s$ per process.
\State \textbf{for} each $\mathcal{D}_s$ \textbf{do} in parallel
\State \hskip0.5em  Assign Adams--Bashforth initial values.
\State \hskip0.5em \textbf{for} each time--step k \textbf{do}
\State \hskip0.75em \textbf{for} each direction $\hth_m$ \textbf{do} \emph{}
\State \hskip1.0em \textbf{for} each $y_j$ \textbf{do} along $x$
\State \hskip1.5em Apply Fourier Continuation to $u_{m}^k(x_i,y_j)$.
\State \hskip1.5em Evaluate $\partial u_{m}^k(x_i,y_j)/\partial x$ using~\eqref{eq:RTEFCSumder}.
\State \hskip1.0em\textbf{end for}
\State \hskip1.0em\textbf{for} each $x_i$ \textbf{do} along y 
\State \hskip1.5em Apply Fourier Continuation to $u_{m,i}^k(x_i,y_j)$.
\State \hskip1.5em Evaluate $\partial u_{m}^k(x_i,y)/\partial y$ using~\eqref{eq:RTEFCSumder}.
\State \hskip1.0em\textbf{end for}
\State \hskip1.0em Evaluate the right hand side of equation \eqref{eq:RTEAB4}.
\State \hskip1.0em Impose boundary conditions.
\State \hskip1.0em Exchange non-physical boundaries with neighboring subdomains.
\State \hskip0.75em\textbf{end for}
\State \hskip0.5em\textbf{end for}

\State  \textbf{end for}
\end{algorithmic}
\end{algorithm}

\section{Numerical Results I: Direct RTE problem}
\label{sec:forwsim}
A Fortran 90 parallel implementation of the proposed algorithm for the
RTE problem~\eqref{eq:RTE}, parallelized with MPI and compiled with
the intel Fortran compiler, was used to produce all of the numerical
results presented in this paper.  In all cases the parallel code was
run on various numbers of nodes on a 16-node cluster, wherein each
node contains an Intel Xeon E5-2630 v3 at 2.40GHz CPU with 24 physical
cores and 128Gb of RAM per node. For geometrical simplicity, only
sixteen cores per node were used, to match a square geometry
containing multiples of $4\times 4$ subdomains.

Three main examples are presented in this section, concerning parallel
scaling, accuracy and a demonstration of a laboratory-type simulation.

\begin{figure}[h!]
\centering
  \includegraphics[width=0.8\linewidth]{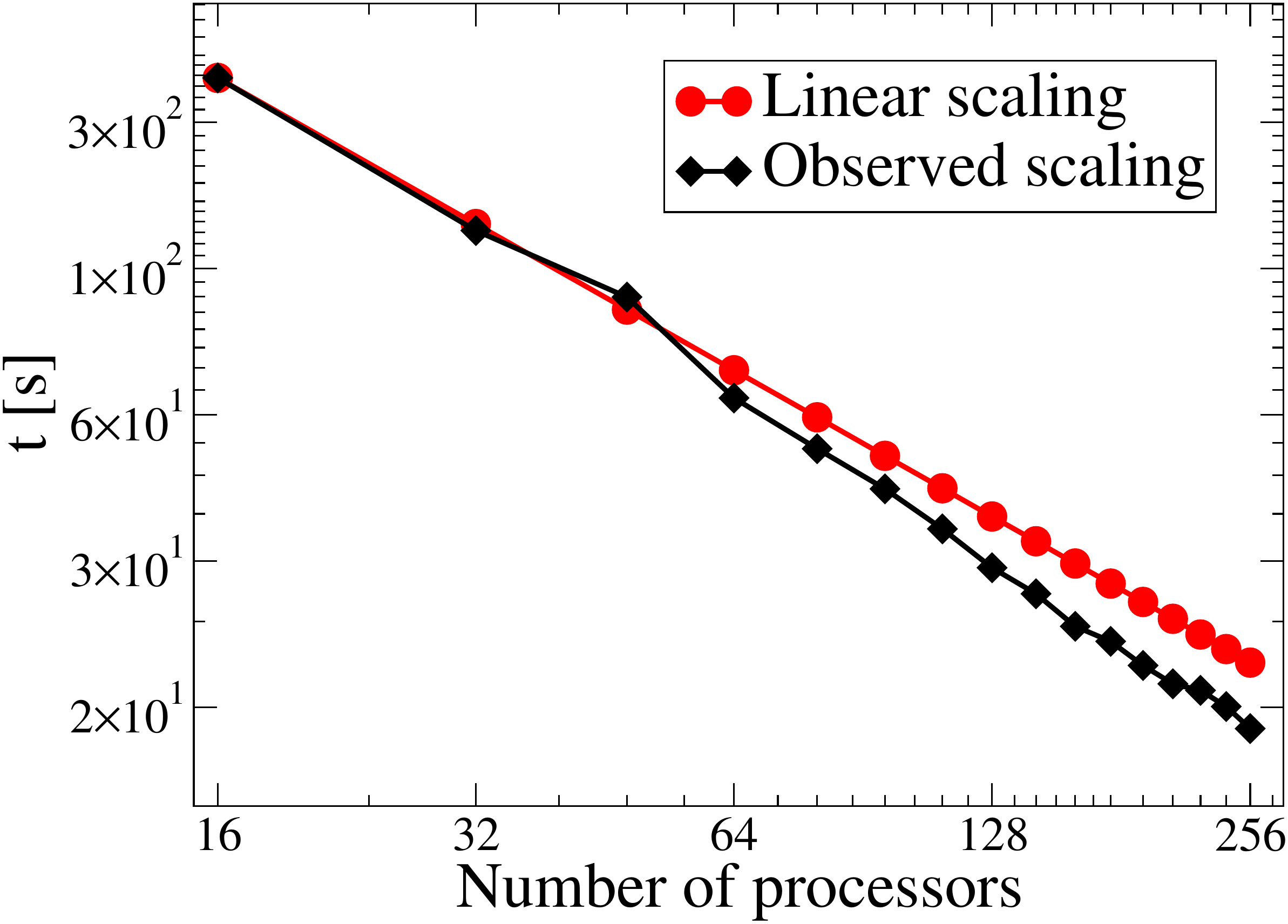}
  \caption{(Color online). Observed scalability of the proposed algorithm 
  for the model problem detailed in the text. 
The computational time on 256 processors, $t_{256}$ for perfect scalability (red circles) produces a speedup $t_{256}/t_{16}=16$ with respect to the computational time for 
the simulation in 16 processors, $t_{16}$. The measured scalability 
with the proposed FC--DOM parallel algorithm (black diamonds) produces a speedup $t_{256}/t_{16}=21.872$, providing an efficiency of $136.7\%$. 
}
 \label{fig:scala}
\end{figure}

Our first test concerns computational cost. Figure~\ref{fig:scala}
presents the computational times required by the proposed algorithm
for a fixed model problem run on various number of processing
cores---a type of test known in the literature as a ``strong scaling
test''.  Constant values of the absorption and scattering coefficients
were used for this test (although, of course, such selections do not
affect the computing time), and the algorithm was run for a
discretization with $N_x=N_y=2000$ and of $M=16$ directions, for a
total of $T=1000$ time steps. Computing times for other final times or
discretization sizes, or even other numbers of computing cores, can
easily be estimated from the results presented in
Figure~\ref{fig:scala} in view of the linear scaling of the method. To
avoid irregularities in the parallel acceleration caused by the
``Intel turbo'' technology, which operates as a few cores are used per
node, but which is ``incrementally'' turned off as processors are
incorporated in a run, we present scaling data up to sixteen nodes,
using a single node as reference, and we report the computing times
observed as the number of nodes increases.  As indicated by the
figure, the proposed algorithm enjoys \textit{perfect parallel
  scaling} (and, even somewhat better than perfect, on account of the
logarithmic cost factors associated with the FFT
algorithm~\cite{Albin2011}) at least up to the 256 processing cores
used---and, we conjecture, for arbitrarily large number of processing
cores, as long as the computational domain can reasonably be
decomposed in a corresponding number of subdomains.

\begin{figure}[h!]
\centering
  \includegraphics[width=0.8\linewidth]{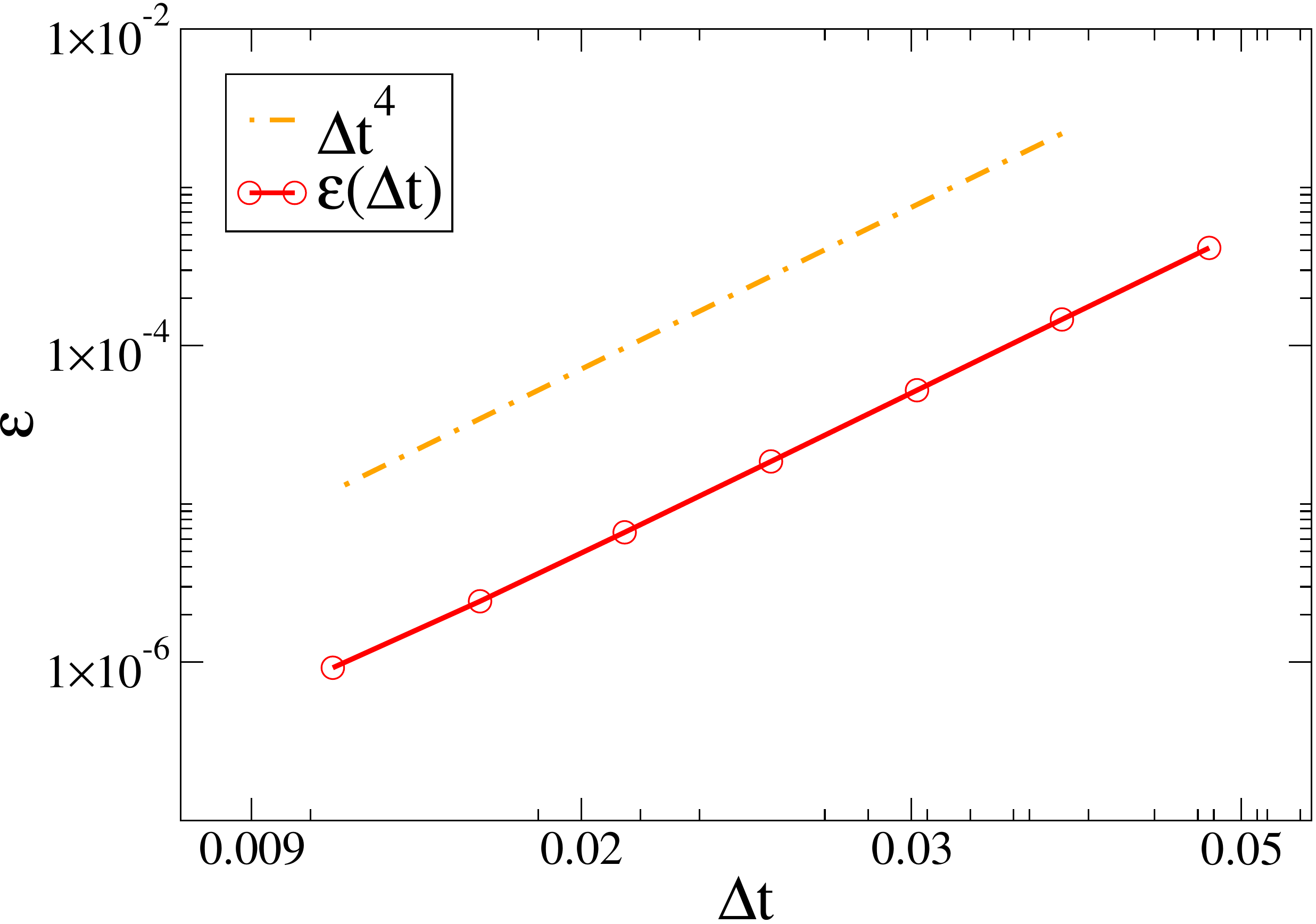}\\
    \vspace{0.3cm}
  \includegraphics[width=0.8\linewidth]{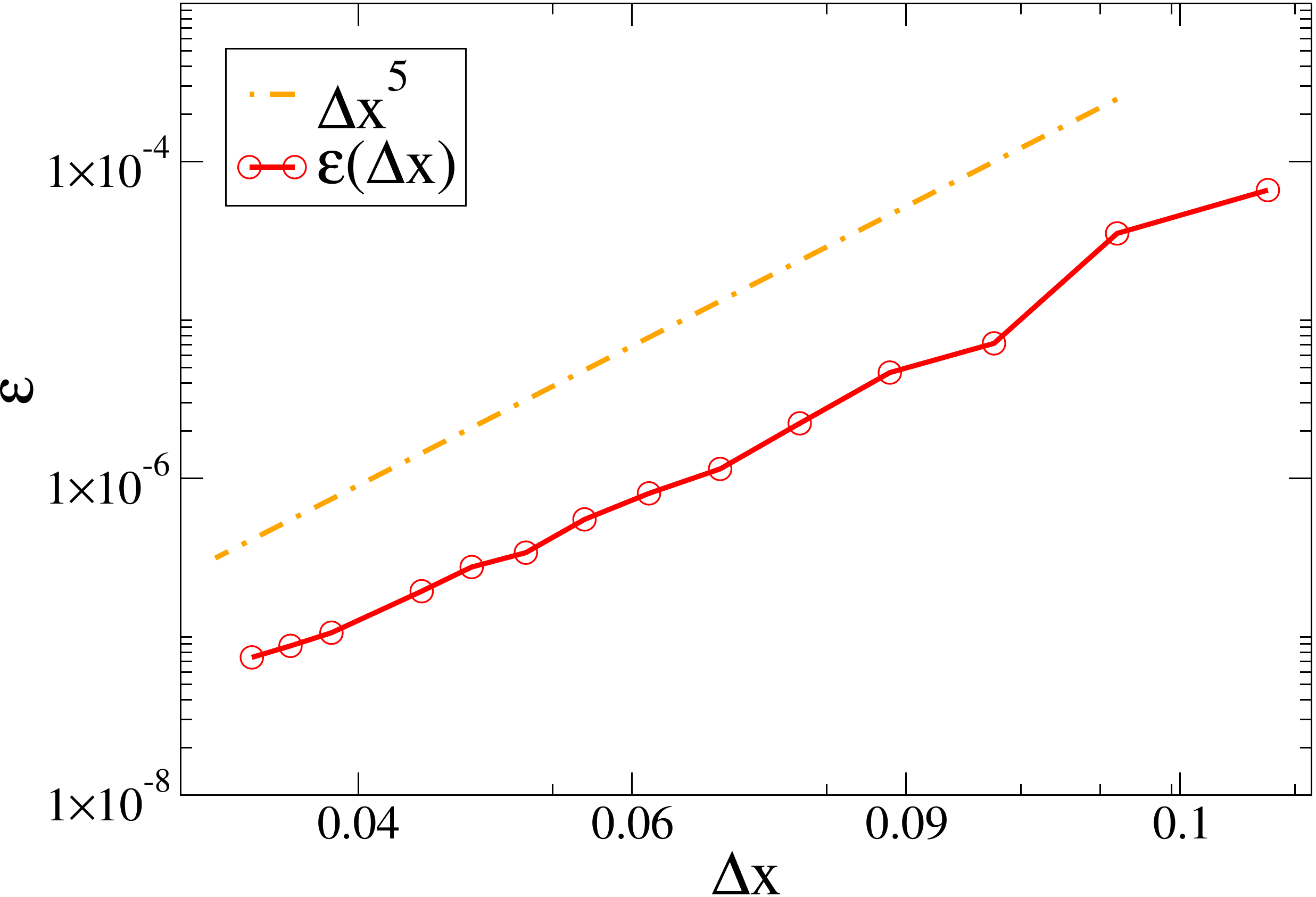}\\
    \vspace{0.3cm}
  \includegraphics[width=0.8\linewidth]{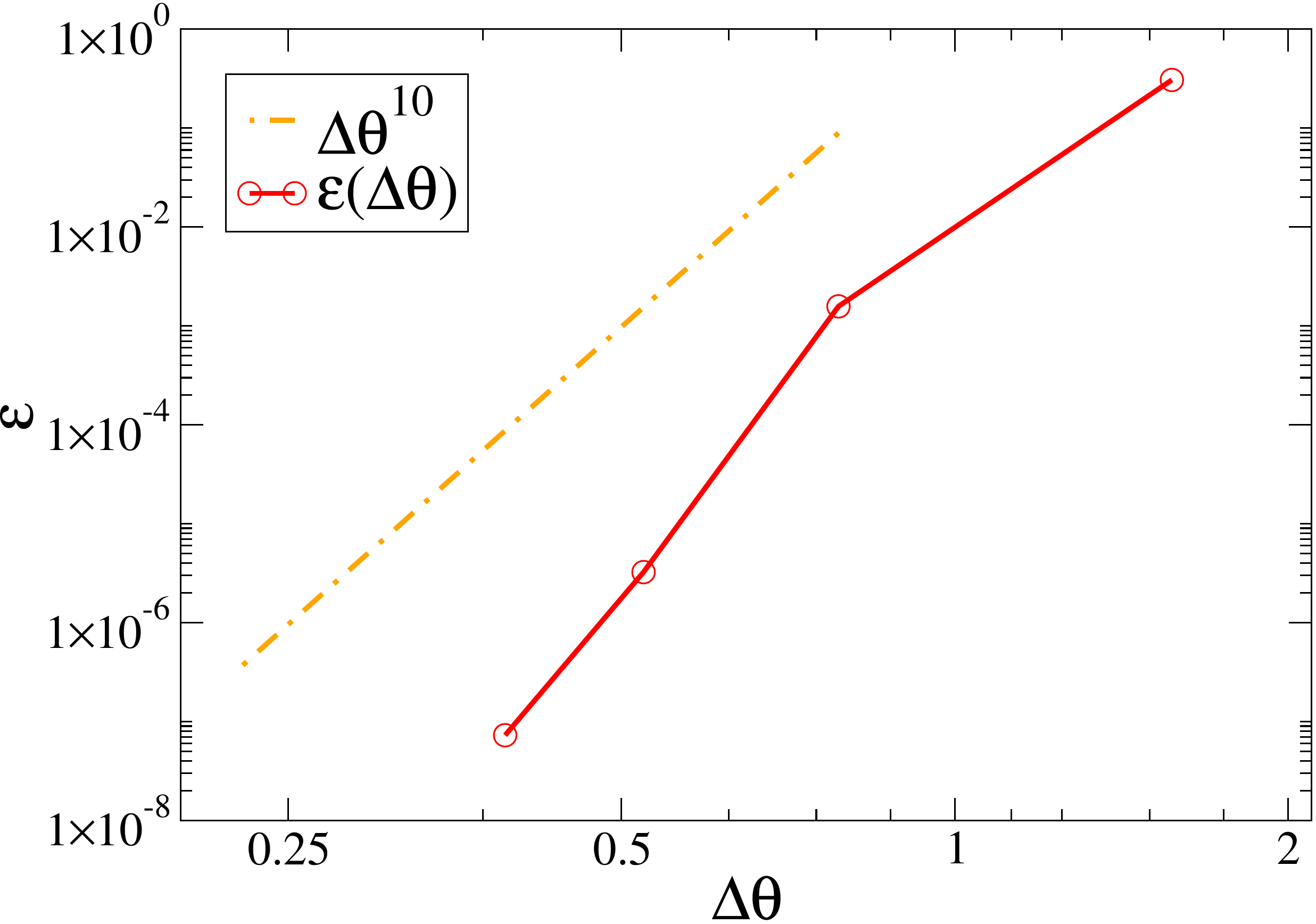}
  \caption{(Color online).  Convergence for the FC--DOM parallel
    solver over each variable. The plot is shown in log-log scale. The
    dash dotted lines shows the slope for the given convergence
    order. The solver shows spectral convergence over the angular
    variable $\theta$.}
 \label{fig:convman}
\end{figure}

The second example in this section concerns the numerical convergence
properties of the algorithm. To study solution errors we consider the
``manufactured-solution'' RTE problem
\begin{equation}
\begin{split}
\begin{aligned}
&\mathcal{T}\big[u[a],a\big]=q,  \; (\x,\theta)  \in \Omega\times [0,2\pi)\\
&u(\x,\theta,t=0)=u_0, \; (\x,\theta)  \in \Omega\times [0,2\pi),  \\
&u(\x,\theta,t)=u_b, \; (\x,\theta) \in \Gamma_-.
\end{aligned}
\end{split}
\label{eq:RTEdisp}
\end{equation}
for which the exact solution
\begin{equation}
u^{\text{an}}(\x,\theta,t)=e^{-(x-t)^2-(y-t)^2-\cos(\theta)^2}.
\label{eq:RTEmansol}
\end{equation}
is prescribed and accounted for by selecting the right-hand side $q$
and initial and boundary conditions $u_0$ and $u_b$ that result as the
proposed solution~\eqref{eq:RTEmansol} is substituted
in~\eqref{eq:RTEdisp}.  For this test we use the square domain
$\Omega=[x_{\text{min}},x_{\text{max}}]\times
[y_{\text{min}},y_{\text{max}}]$, with
$x_{\text{min}}=y_{\text{min}}=0$ cm and
$x_{\text{max}}=y_{\text{max}}=3$ cm, and we evolve the solution in
the time interval $0\leq t\leq T=3$ ps. We use a domain decomposition
with eight subdomains. The scattering medium considered is isotropic
and homogeneous, with $g=0$, $a=0.35/$cm and $b=20/$cm. We study the
convergence properties on all the variables involved, evaluating
maximum errors, in each case, by comparison with numerical scalar
fluxes~\eqref{eq:photondensity} against the analytic scalar flux
\begin{equation*}
\begin{split}
\begin{aligned}
  \Phi^{\text{an}}(\x,t)&=\int_{2\pi} u^{an}(\x,\theta,t) d\theta=
  k_{\Phi}\times e^{-(x-t)^2-(y-t)^2},\\
  k_{\Phi}&=\int_{2\pi}e^{-\cos(\theta)^2}d\theta\simeq
  4.052876133898710,
\end{aligned}
\end{split}
\label{eq:photondensityan}
\end{equation*}
where $k_{\Phi}$ was obtained, with 16-digit accuracy, using the
software Wolfram Mathematica. The solution was evolved up to a fixed
final time $T$, and the maximum error was then evaluated by means of
the expression
\begin{equation*}
\varepsilon=\text{max}_{\x \in \Omega} |\Phi^{\text{num}}(\x,T) -\Phi^{\text{an}}(\x,T)|.
\label{eq:maximumerror}
\end{equation*}
The high order convergence of the FC--DOM parallel approach against
the proposed manufactured solution is demonstrated in
figure~\ref{fig:convman}, which displays convergence curves as
$\Delta t$, $\Delta x=\Delta y$ and $\Delta\theta$ are refined. (When
considering refinements in one of the variables, the mesh sizes in the
other variables were kept fixed at sufficiently fine levels, so as to
avoid error cross-contamination.) Clearly, excellent convergence is
observed in all three cases.
\begin{figure*}[h!]
\centering
  \includegraphics[width=0.325\textwidth]{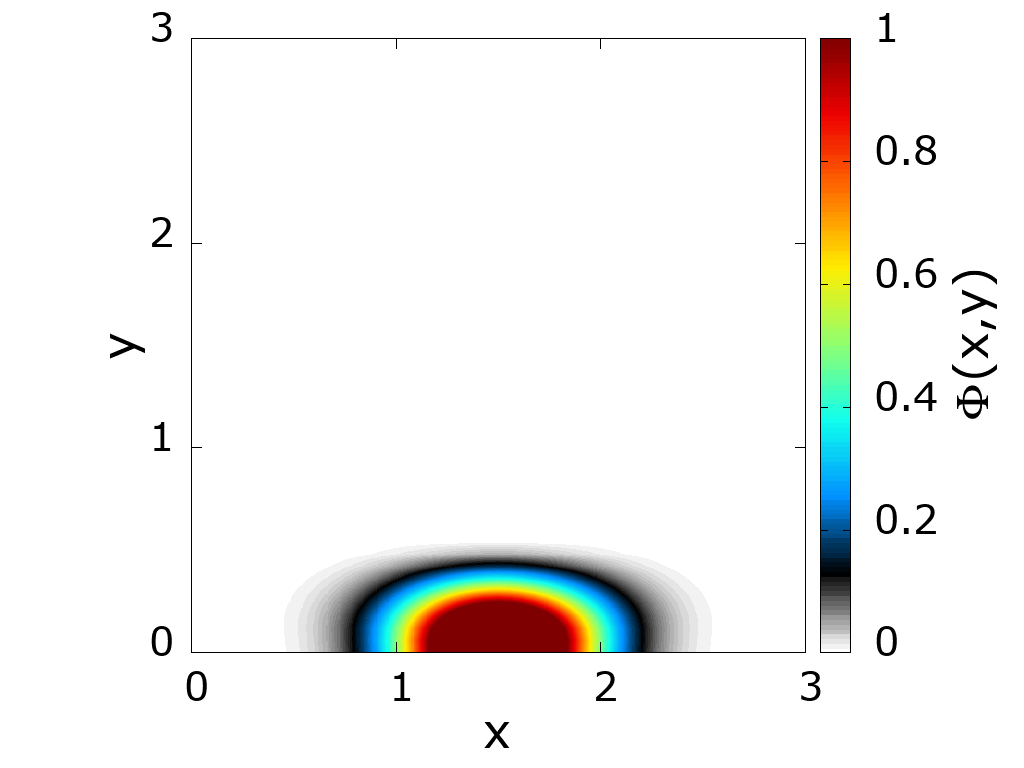}
  \includegraphics[width=0.325\textwidth]{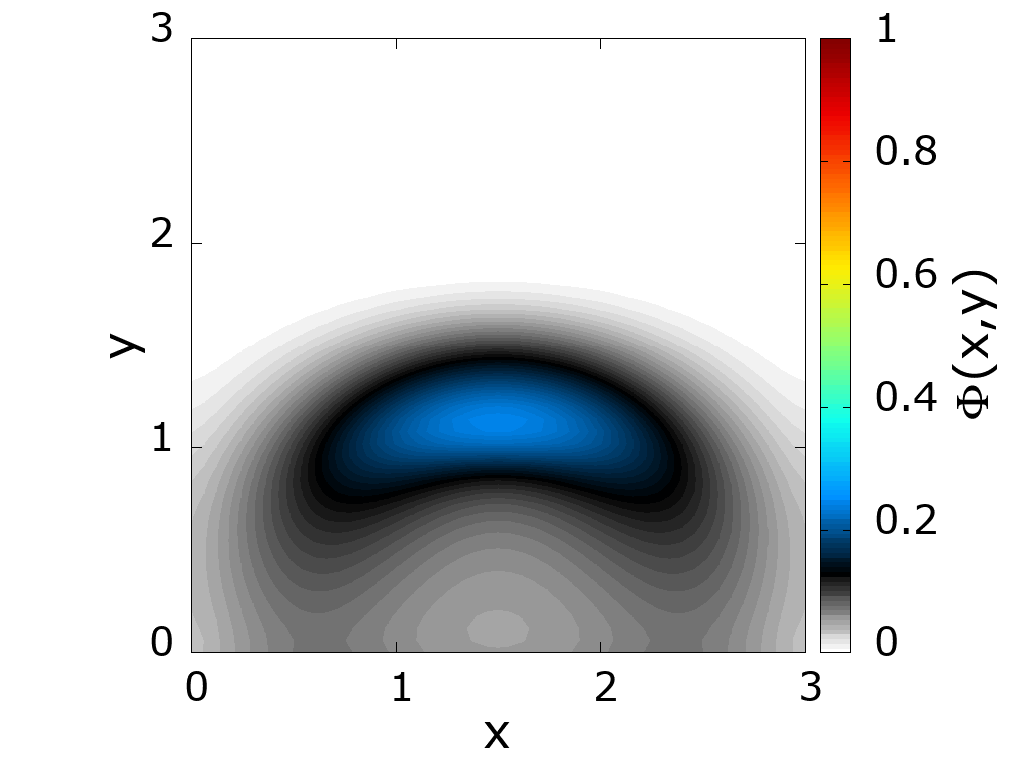}
  \includegraphics[width=0.325\textwidth]{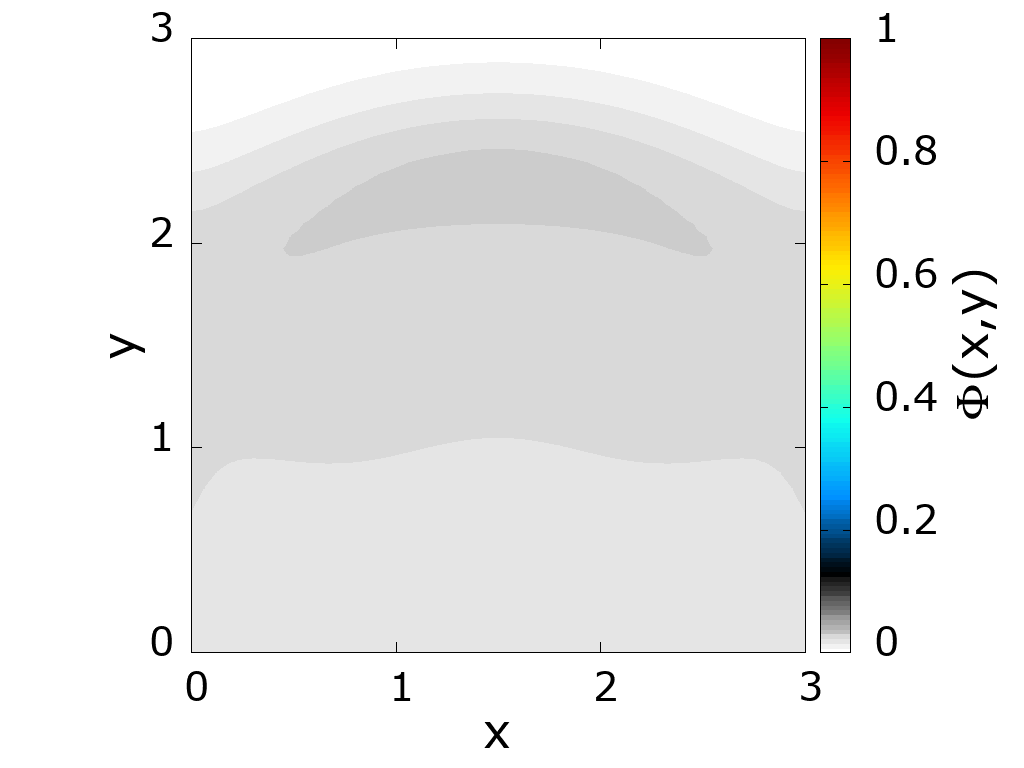}
  \caption{(Color online).  Photon transport simulation produced by
    the parallel FC-DOM algorithm presented above. The scalar flux
    eq.~\eqref{eq:photondensity} is shown at three different times
    from left to right, $t=30$ps, $t=100$ps and $t=170$ps, for a
    source injecting radiation at $\x_s=(1.5,0)$cm.}
 \label{fig:photonflux}
\end{figure*}

In actual optical tomography contexts, for which an analytic solution
is, of course, not known, typically collimated irradiation from a
pulsed laser beam is incident at the boundary of the domain; the third
and last example in this section, which is presented in what follows,
concerns precisely such a scenario. In our test case the collimated
irradiation from laser pulses is modeled by means of a peaked function
in the directional variable, $\theta$, wherein the function peak
coincides with the laser beam direction.  We solve equation
\eqref{eq:RTE} with Fresnel boundary conditions and using the
well-known approach based on use of a Gaussian spatial variation for a
collimated laser-beam source:
\begin{equation}
\begin{aligned}
\begin{split}
  q(\x,\theta,t)=&\exp(-|\x-\x_s|^2/
  2\sigma_\x^2)\times w\left(\frac{|\theta-\theta_s|}{\sigma_{\theta}}\right)\\
  & \times w\left(\frac{|t-t_s|}{\sigma_t}\right);
\end{split}
\end{aligned}
\label{eq:sourcelab}
\end{equation}
the beam dependence on the temporal and angular variables, in turn,
are modeled on the basis of the window
function~\eqref{eq:sourcelabwind}, but any other models of these
variations could be used without difficulty. For our example the laser
source is located at $\x_s=(1.5,0)$ with spatial spread
$\sigma_\x=0.3$cm, it points in direction $\theta_s=\pi/2$ with
angular spread $\sigma_{\theta}=\pi/4$, and is temporally centered at
$t_s=0$, with temporal spread $\sigma_t=30$ ps. By choosing a smooth
source, we avoid the appearance of singularities in the source
function that would be inherited by the RTE solution and deteriorate
the convergence properties of the method. We solve the
RTE~\eqref{eq:RTE} in the 2D spatial domain
$\Omega=[x_{\text{min}},x_{\text{max}}]\times[y_{\text{min}},y_{\text{max}}]=[0,3]\times[0,3]$
with values $a(\x)=0.35$/cm, $b(\x)=20$/cm, $g=0.8$, $n_{\Omega}=1.4$
and $n_0=1$. Figure~(\ref{fig:photonflux}) displays the scalar flux
eq.~\eqref{eq:photondensity} obtained for this problem; clearly a
smooth spatio-temporal distribution is obtained suggesting a solution
of high quality in accordance with the accuracy studies presented
above in this section.

\section{Inverse problem solver}
\label{sec:inversesolv}

As indicated in Section~\ref{sec:introduction}, the inverse problem
solver proposed in this paper incorporates, in particular, the novel
MSS strategy based on use of multiple staggered sources instead of the
transport sweep TS strategy that underlies the previous related
literature. In the previous TS source method, a separate
forward-adjoint pair of simulations is used for each one of the
laser-source illuminations, and the results are then combined to
achieve the inversion via gradient descent. The proposed MSS approach,
in turn, constructs the objective function on the basis of
time-staggered laser sources in conjunction with {\em a single pair of
  forward and adjoint simulations}, and thus, as demonstrated in
Section~\ref{sec:inverseres} (quantitatively in
Figures~\ref{fig:itdet} and~\ref{fig:ithead}, and qualitatively in
Figures~\ref{fig:reconst} and~\ref{fig:rechead}), it significantly
reduces the computing time required for the solution of the inverse
problem (e.g.  by a factor of six in Figure~\ref{fig:ithead}) without
any deterioration in image quality.
\begin{figure}[h!]
\centering
  \includegraphics[width=0.8\linewidth]{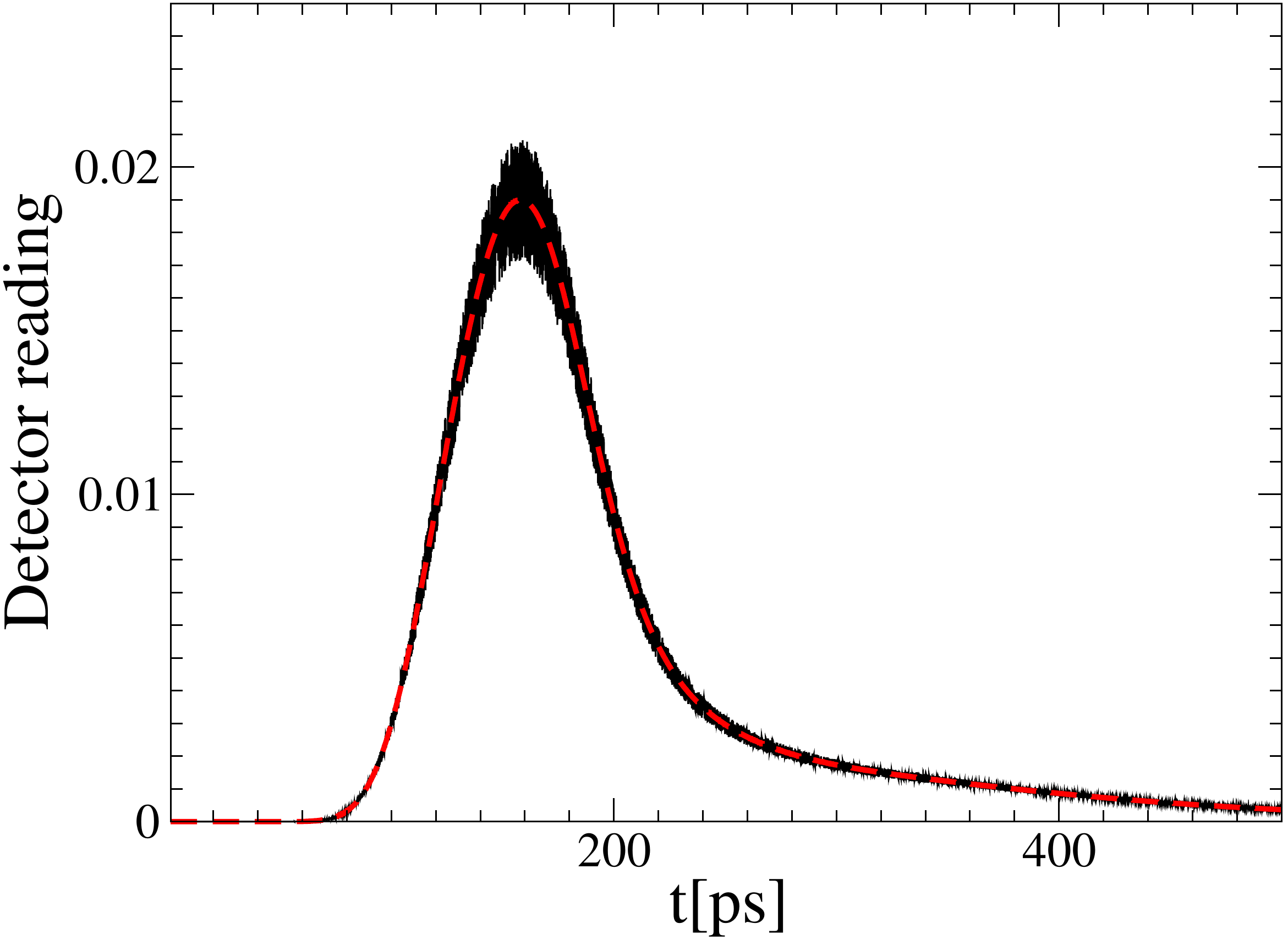}\\
    \vspace{0.3cm}
  \includegraphics[width=0.8\linewidth]{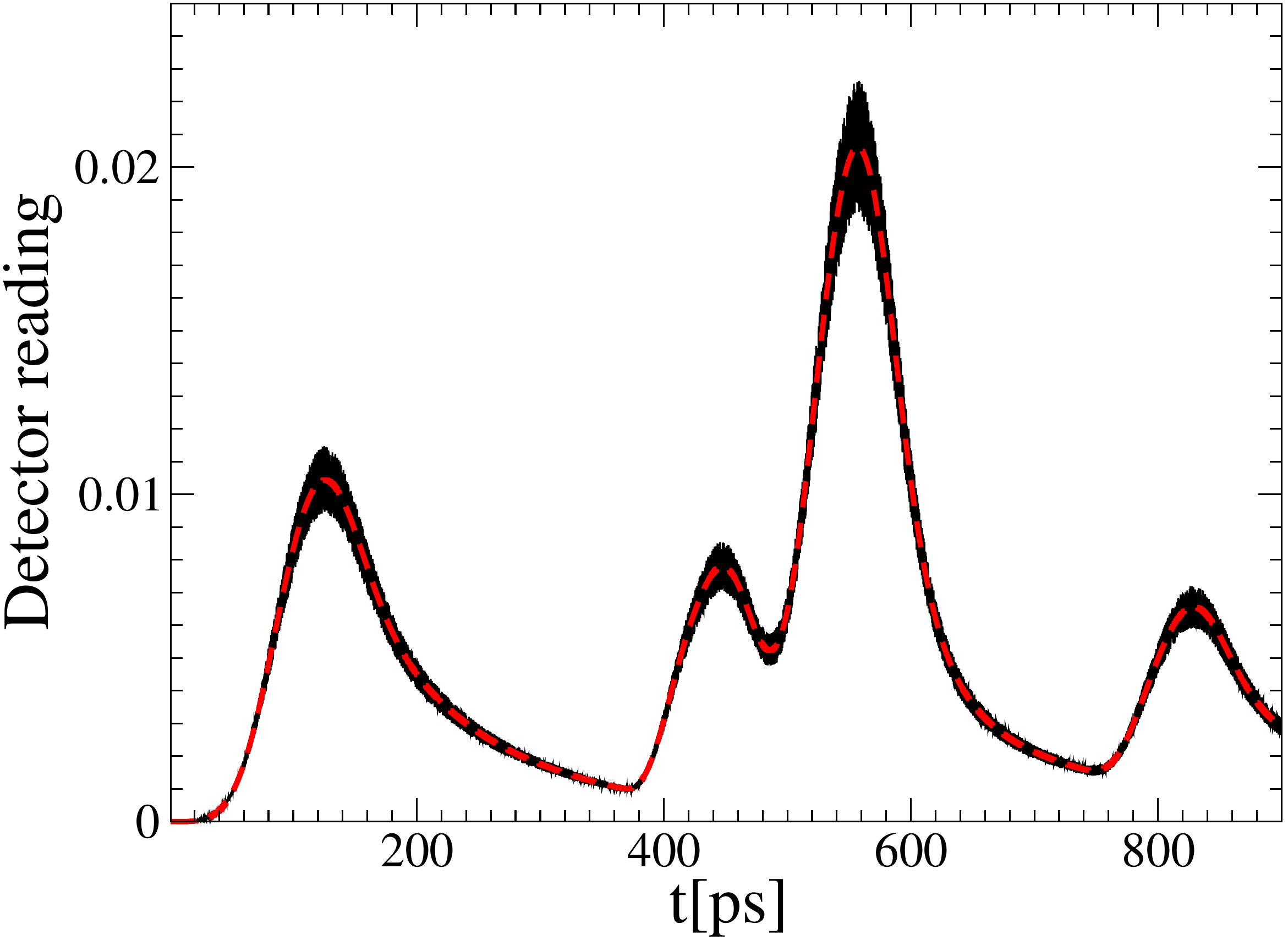}
  \caption{Detector readings ($G_1[u_1]$ in eq.~\eqref{eq:OpMed}) for
    the TS and MSS methods, with $N_s=1$ source for TS method, and
    $N_s=4$ sources for the MSS method, and with and without addition
    of 10\% of noise to the signal. The detector was placed at
    $\x_d=(3.0,2.25)$. Each single MSS laser beam is located at the
    center of a face of the square domain, and the activation times
    $\tau_{1,1}=0$ps, $\tau_{2,1}=200$ps, $\tau_{3,1}=400$ps, and
    $\tau_{4,1}=600$ps were used. Clearly, the MSS detector readings
    combine signals arising from all sources. }
 \label{fig:detread}
\end{figure}

Detector readings corresponding to a TS source and an MSS source are
displayed in the upper and lower panels of Figure~\ref{fig:detread},
respectively; both noiseless and noisy data with $10\%$ random noise
added, are presented in the figure. The MSS time delays introduce a
degree of decoupling in the portions of the detector signals
originating from laser beams applied at different locations, thus
enabling effective inversions without requiring independent forward
and backward solutions for each one of the beams separately. As
indicated in Section~\ref{sec:formulat}, depending on parameter
choices, the expression~\eqref{eq:RTEsources} yields sources
corresponding to the TS or MSS illumination approach, and
Section~\ref{sec:inverseres} presents results obtained from both the
TS and MSS methods.

We solve both the TS and MSS inverse problems on the basis of the
iterative lm-BFGS~\cite{Byrd1995} quasi-Newton gradient descent
method. In detail, the inverse solver seeks the absorption coefficient
function that minimizes eq.~\eqref{eq:FObjpr} subject to the
constraints $a^{\ell} \leq a(\x) \leq a^u$, i.e.  we seek the
absorption function $a = a(\x)$ given by
\begin{equation}
  \begin{split}
\begin{aligned}
  a = \text{argmin}_{a^{\ell} \leq \widetilde a(\x) \leq a^u} \Lambda[\widetilde a].
\end{aligned}
\end{split}
\label{eq:minprob}
\end{equation}
The constraints on the values of admissible functions
$\widetilde a(\x)$ are known as prior knowledge on the general
absorption properties of the tissue under consideration.

The inverse solver, which is summarized in Algorithm~\ref{algfcinv},
proceeds as follows. Starting from an initial guess $a^0(\x)$ for the
absorption coefficient and using the experimental detector readings
$\widetilde{G}_{j,i}$, with $j=1, \ldots,N_d$ and $i=1,\ldots,N_q$,
the forward and adjoint problems~\eqref{eq:RTE}
and~\eqref{eq:AdjointProblem} are solved, and the
gradient~\eqref{eq:FDwin22} is computed. The functional gradient is
then passed to the lm-BFGS algorithm, which returns an updated
absorption coefficient $a^1(\x)$ that reduces the mismatches between
the experimental and simulated detector readings. The procedure is
iterated with lm-BFGS convergence to a minimum of the objective
function~\eqref{eq:FObjpr}.

\begin{algorithm}
\caption{Parallel Inverse Problem Solver}\label{algfcinv}
\begin{algorithmic}[1]
\State \textbf{for} iteration $i=1,\ldots,i_{\text{max}}$ \textbf{do}
\State \hskip0.5em \textbf{for} each generalized source $q_j, j = 1,\ldots,N_q$ \textbf{do}
\State \hskip0.75em  solve the forward problem by means of algorithm \ref{algfc}
\State \hskip0.75em solve the adjoint problem by means of algorithm \ref{algfc}
\State \hskip0.5em\textbf{end for}
\State \hskip0.5em Construct the gradient eq.~\eqref{eq:FDwin22}, call lm-BFGS and update the coefficient $a^i(\x)$.       
\State  \textbf{end for}
\end{algorithmic}
\end{algorithm}

\section{Numerical Results II: Inverse problem}
\label{sec:inverseres}

This section demonstrates the character of the proposed inverse
problem solver for two main model problems, namely, 1)~Imaging of
cancerous tissue within a human neck section; and, 2)~Hemodynamic
response in a human head model. In both cases, the optical tomography
inverse problems under consideration concerns configurations in which
inclusions characterized by absorption higher than that of the
surrounding tissue are to be imaged; the higher absorption values
arise from the excess of oxygenated hemoglobin under the presence of a
tumor in tissue (due to the tumoral angiogenesis) in the neck-tumor
problem~\cite{Zhu2005,Zhu2010,Fujii2016b,Althobaiti2017,Guven2003},
and from radiation absorption excess originated by the presence of
oxygenated hemoglobin triggered by hemodynamic response due to the
activation of a brain
region~\cite{Boas2001,bluestone2001,Arridge1999,Hernandez2020}, in the
brain-imaging problem. Accordingly, in what follows the lower bound
$a^{\ell}(\x)$ for $a(\x)$ in~\eqref{eq:minprob} is set to equal the
absorption value of the background tissue (which is assumed to be
known a priori), and the corresponding upper bound is set to
$a^u(\x)=1$/cm, which provides a reasonable upper constraint for the
absorption values of the types of tissue under consideration~\cite{Dehaes2011}.

We solve these inverse problems on the basis of synthetic data
obtained by running the forward problem for a given ``target''
absorption coefficient $a^t(\x)$. In order to account for experimental
noise, we add a 10\% of random noise to the resulting detector
readings prior to the inversion process, as illustrated in
Figure~\ref{fig:detread}.  In particular, we study the convergence of
Algorithm~\ref{algfcinv} for a varying number of sources and detectors
for the given configuration. In order to evaluate the convergence of
the reconstruction process we use the $L^2$-error norm
\begin{equation}
\begin{split}
\begin{aligned}
E(i)=\sqrt{\frac{\int_{\Omega}(a^t(\x)-a^i(\x))^2  d\x}{\int_{\Omega}(a^t(\x))^2 d\x}}
\end{aligned}
\end{split}
\label{eq:errl2}
\end{equation}
where $E(i)$ corresponds to the $L^2$-error in the absorption value
$a^i(\x)$ obtained at the $i$-th iteration of the inverse solver.

\subsection{Neck tumor imaging}
\label{sec:neck}
\begin{figure}[h!]
\centering
  \includegraphics[width=0.4\linewidth]{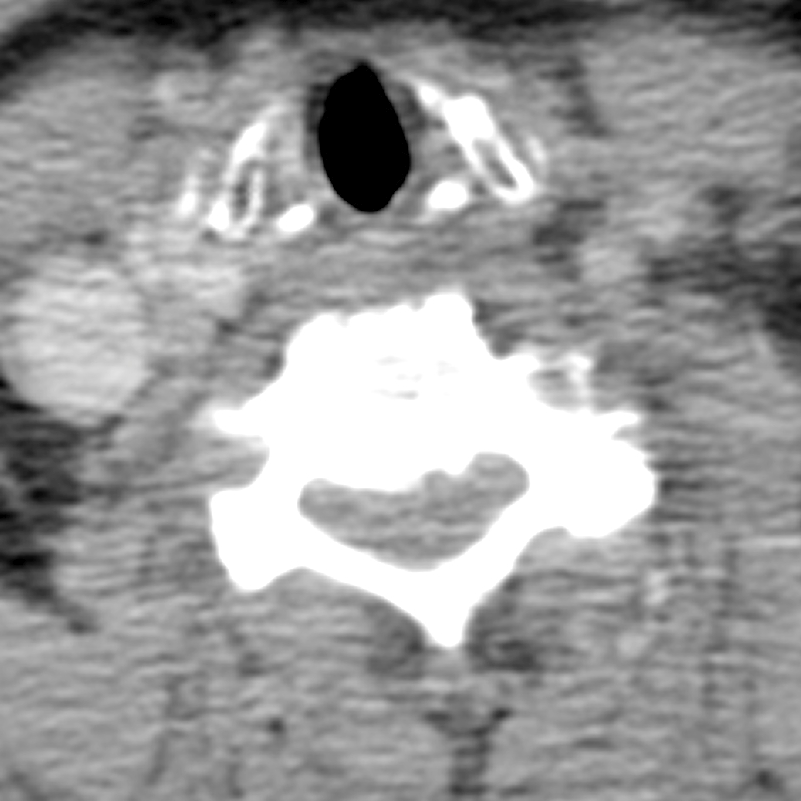}\\
  \caption{Magnetic Resonance Image \cite{CCommons} for the neck model
    employed. The absorption and scattering coefficients were set in
    accordance with reference \cite{Fujii2016}.}
 \label{fig:mriim}
\end{figure}

Our first test case concerns the application of optical tomography for
diagnoses in patients for which a background Magnetic Resonance
Imaging of a neck section is available (Figure~\ref{fig:mriim}); such
situations arise as e.g. imaging of either evolving tumors or new
metastatic tumors is sought within a body part (the neck, in this
case) for which an existing MRI image was acquired months or years in
advance.  The portability and low cost of optical tomographic systems
make optical tomographic devices much more accessible than MRI systems
for such periodic monitoring and diagnostic applications.  Studies for
the forward modeling of light propagation in the human neck are
reported in~\cite{Fujii2016b,Fujii2016}. In the present test case we
consider the inverse problem for such a configuration (which, of
course, requires the repeated solution of forward and adjoint
problems, as described in Algorithm~\ref{algfcinv}), focusing on the
reconstruction itself and illustrating the convergence character and
computational time required by the proposed algorithm.
\begin{figure*}[h!]
\centering
  \includegraphics[width=0.4\textwidth]{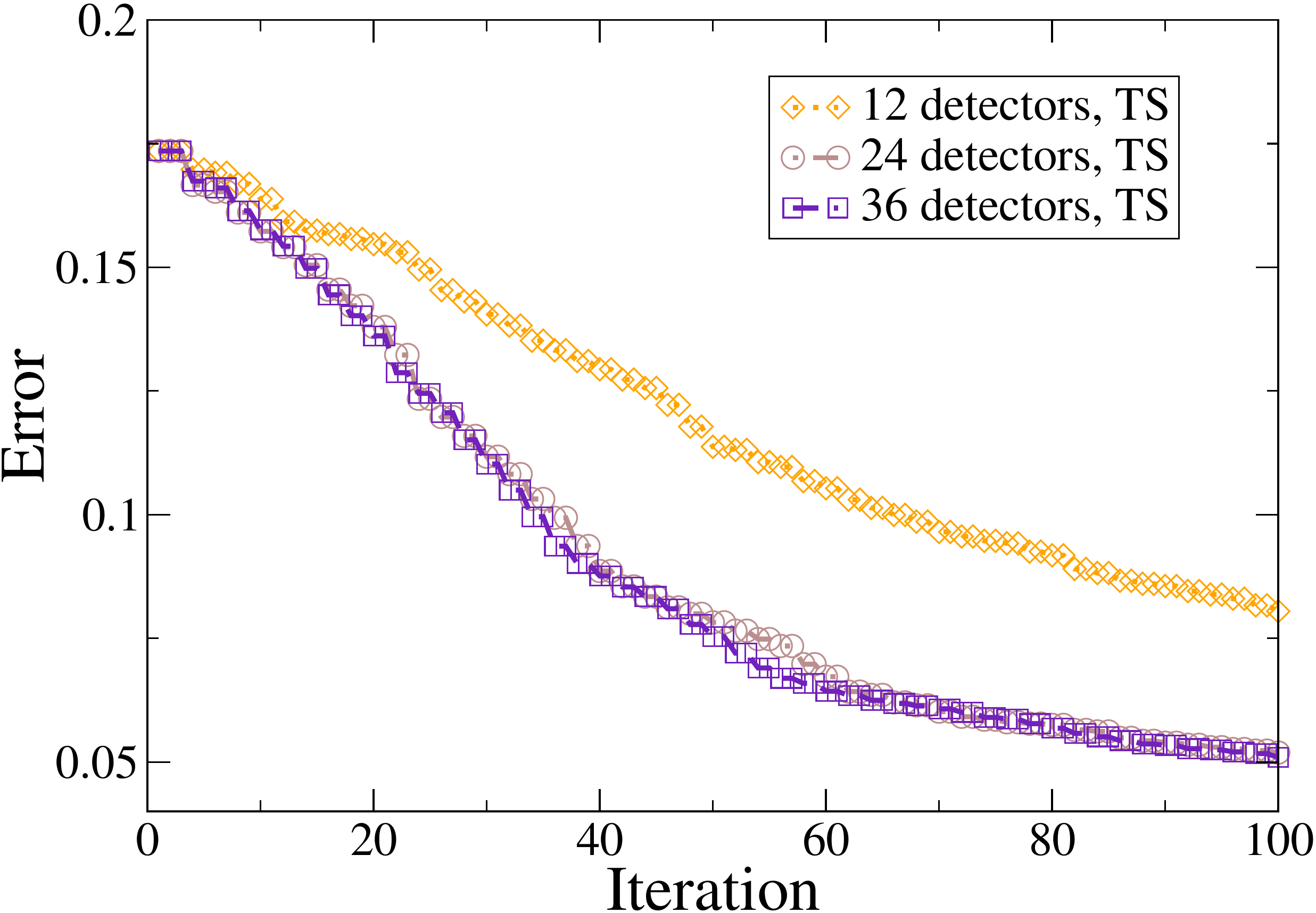}
  \includegraphics[width=0.4\textwidth]{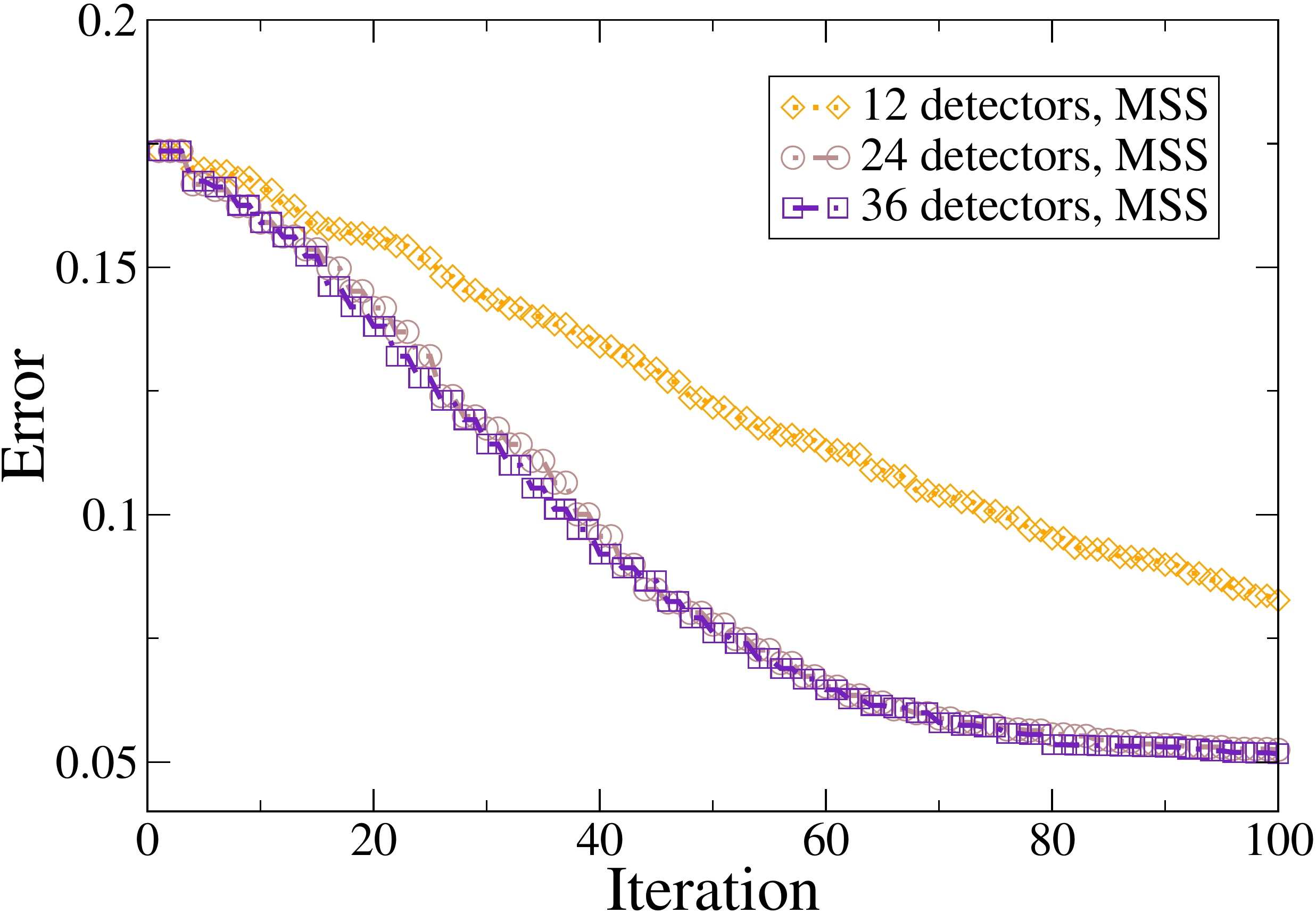}
\\
    \vspace{0.3cm}
  \includegraphics[width=0.4\textwidth]{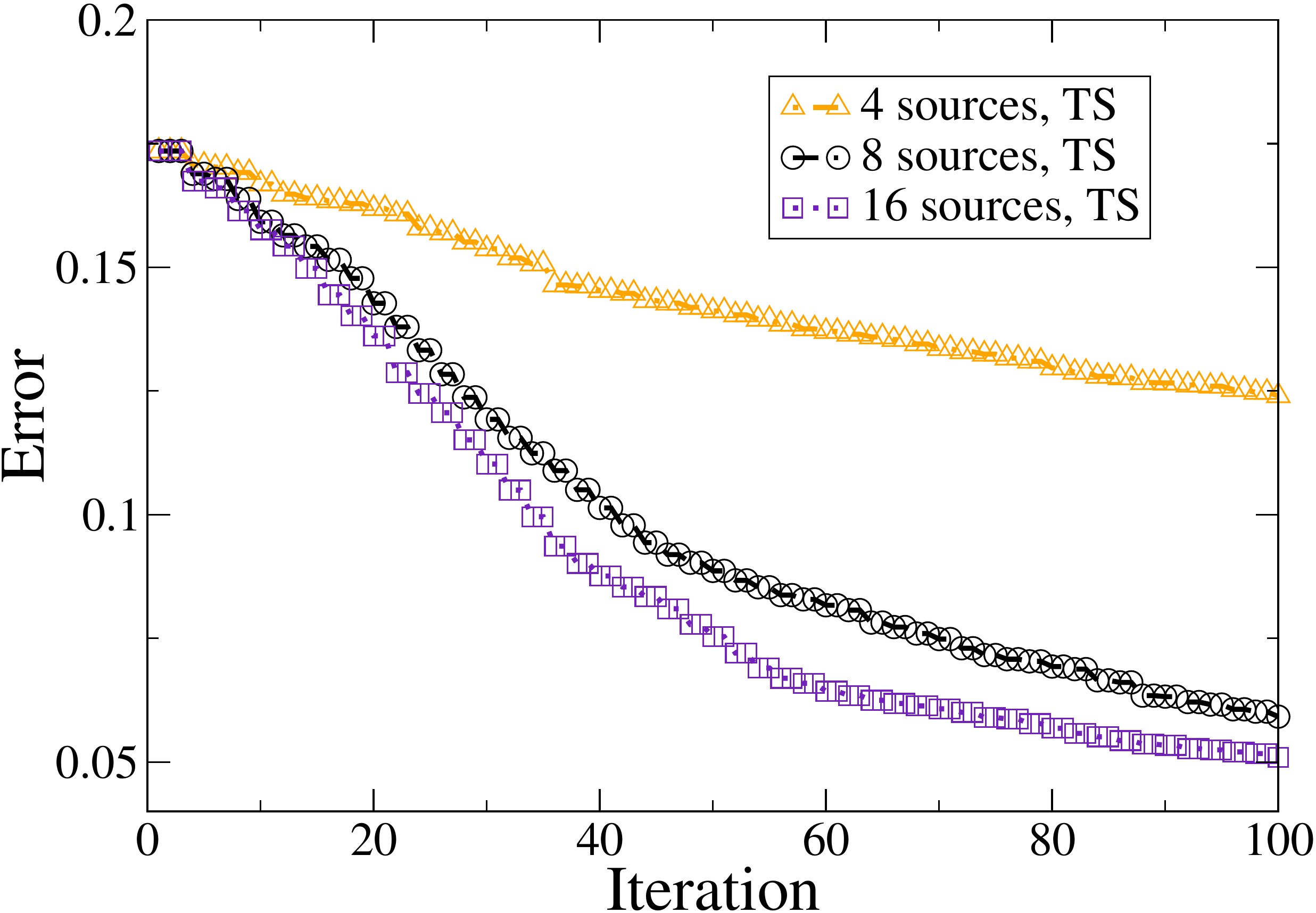}
  \includegraphics[width=0.4\textwidth]{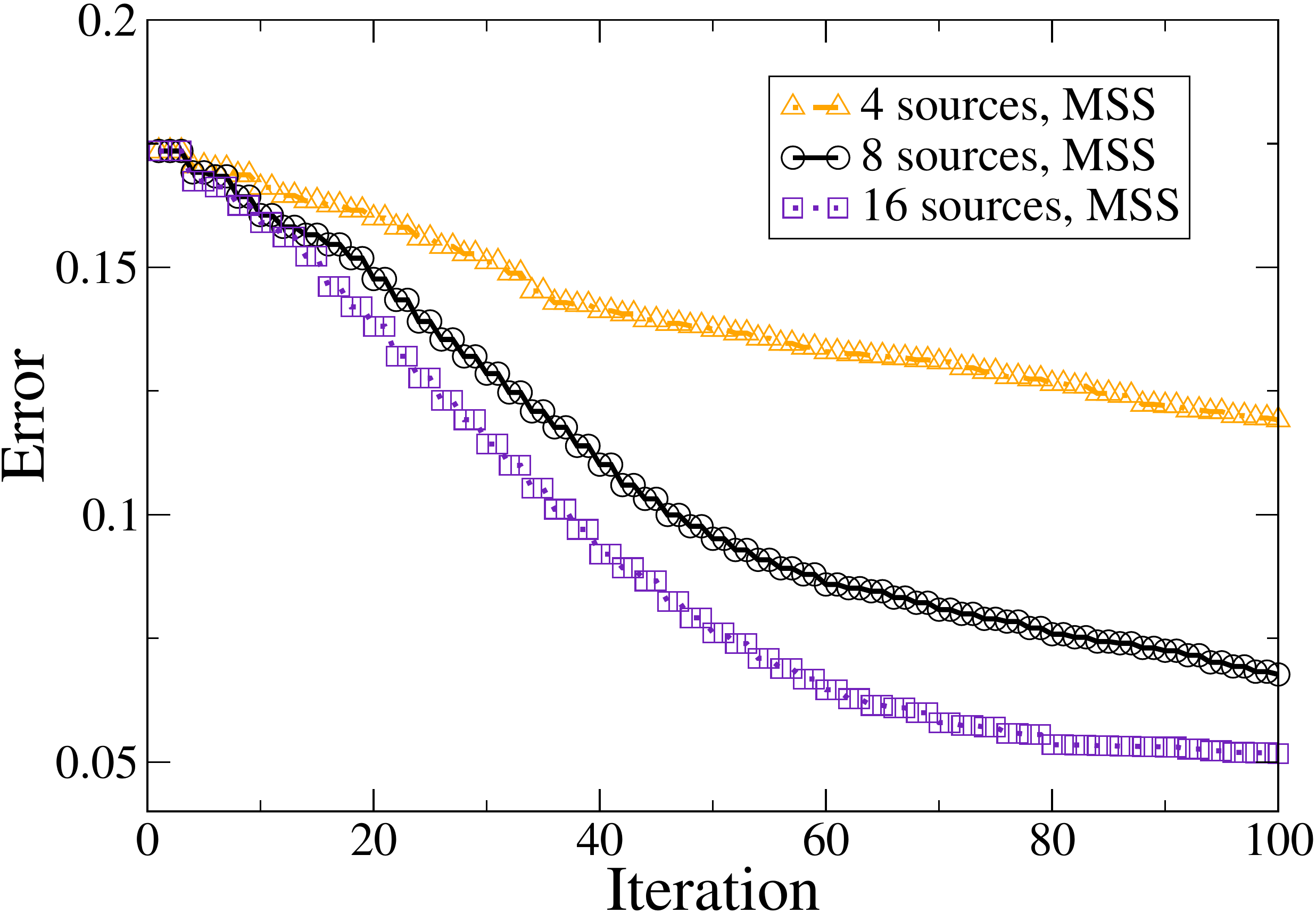}
  \caption{$L^2$-error norm eq.~\eqref{eq:errl2} convergence for the absorption coefficient 
  reconstruction with respect to the number of iterations, for 
  different number of detectors (top) and sources (bottom), for the TS and MSS methods. 
   Left panel: 
  convergence eq.~\eqref{eq:errl2} up to 100 iterations for the TS method. Right panel: convergence eq.~\eqref{eq:errl2} up to 100 iterations for the MSS method. 
  For the simulations in the top panel 16 laser sources were used. For the simulations in the bottom panel, 36 detectors were used.}
 \label{fig:itdet}
\end{figure*}
For the present test case the absorption coefficients for the spine,
the spinal chord, and the trachea will remain fixed in the
reconstruction process (in accordance with the values provided
in~\cite{Fujii2016}), since tumoral angiogenesis is only expected to
exist in the soft tissue. Additionally, we restrict the presence of
tumor inclusions to regions slightly away from the neck boundary, so
as to avoid the error amplification associated with the existence of
exponential boundary layers~\cite{Gaggioli2021} (a full treatment of
such near-boundary imaging configurations is left for future
work). Accordingly, the value of the absorption coefficient in the
proximity of the boundary is set to the background tissue value
$a(\x)=a_b(\x)$ at all points closer than $0.5$cm from the boundary.
\begin{figure*}[h!]
\centering
  \includegraphics[width=0.325\textwidth]{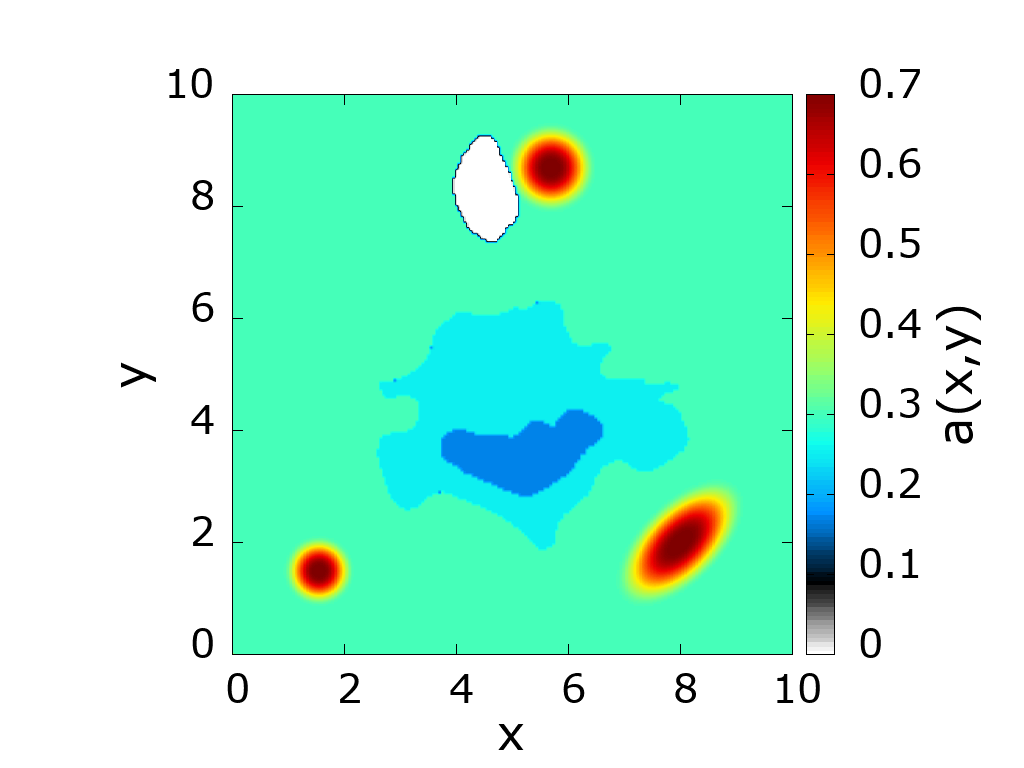}
  \includegraphics[width=0.325\textwidth]{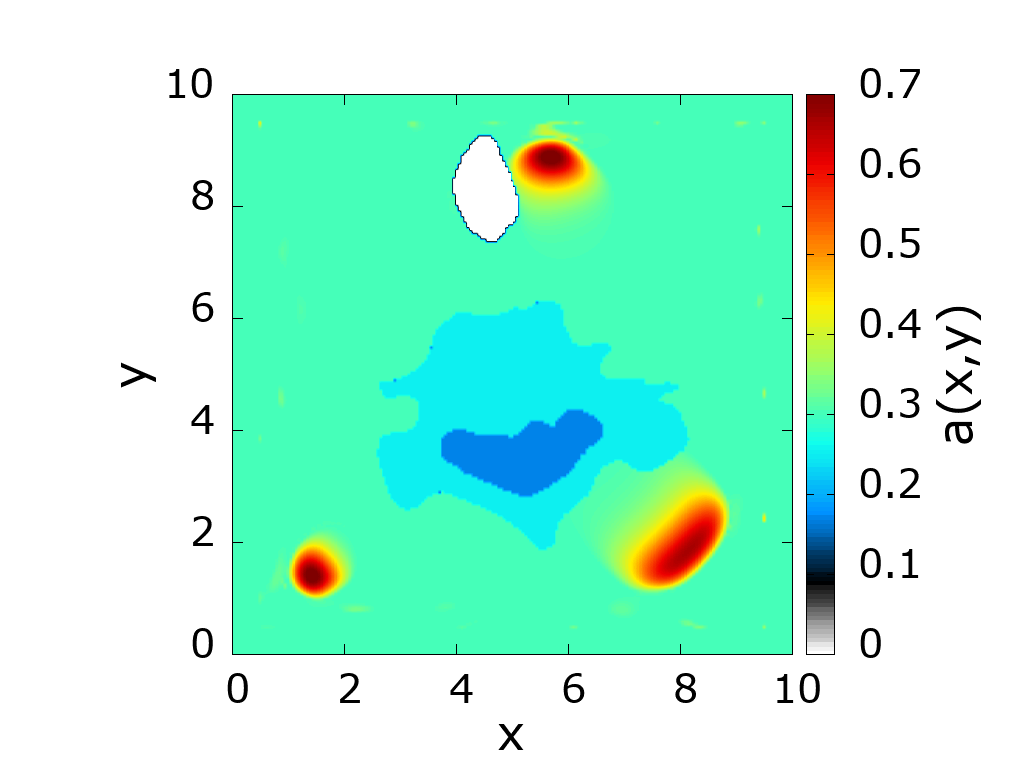}
  \includegraphics[width=0.325\textwidth]{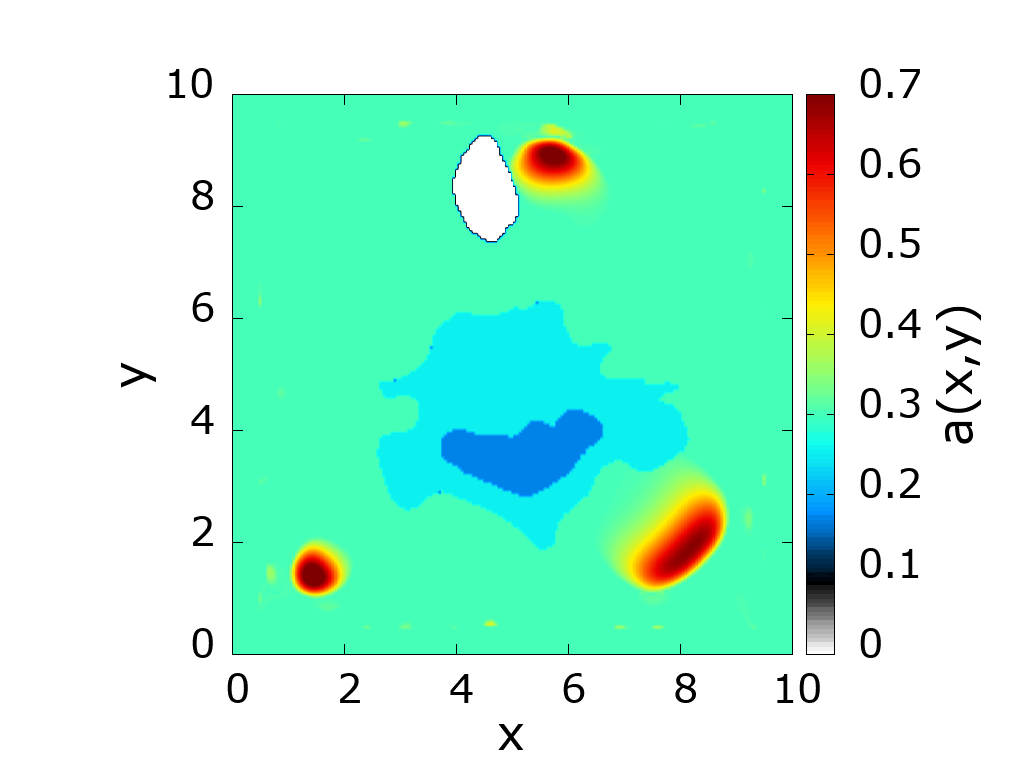}
  \caption{(Color online). From left to right: 
  true absorption coefficient, reconstructed absorption coefficient after 100 iterations 
  using the TS method, and  reconstructed absorption coefficient after 100 iterations
  using the MSS method. For these reconstructions 36 detectors 
  and 16 sources were used.}
 \label{fig:reconst}
\end{figure*}

We tackle the present neck-imaging problem by means of both the TS and
MSS methods described in Section~\ref{sec:formulat}, in a
configuration containing $N_b$ laser beams (see eq.~\eqref{n_beams}),
and we consider examples with $N_b=4$, $8$ and $16$. Per the
description in that section, a single generalized source containing
$N_b = N_s$ beams is utilized in the MSS approach, with $N_q=1$, and
groups of four sources, one per face in the square domain depicted in
Figure~\ref{fig:mriim}, are simultaneously activated, with time delay
of $300$ps between groups of four sources. For example, in the case
$N_b=16$ and letting $\x_1^i$ denote the $i$-th beam in the group of
four beams that is activated first we have $\tau_{1,1}=0$ps for the
simultaneous sources placed at $\x_1^1=(2.0,0.0)$,
$\x_1^2=(10.0,2.0)$, $\x_1^3=(8.0,10.0)$, $\x_1^4=(0.0,8.0)$.  The
remaining sources, located at point $\x_k^i$, activated at time
$\tau_{k,1}$ ($k=2,3,4$) are arranged as follows: $\x_2^5=(4.0,0.0)$,
$\x_2^6=(10.0,4.0)$, $\x_2^7=(6.0,10.0)$, $\x_2^8=(0.0,6.0)$ activated
at $\tau_{2,1}=300$ps; $\x_3^9=(6.0,0.0)$, $\x_3^{10}=(10.0,6.0)$,
$\x_3^{11}=(4.0,10.0)$, $\x_3^{12}=(0.0,4.0)$ activated at
$\tau_{3,1}=600$ps; and $\x_4^{13}=(4.0,0.0)$, $\x_4^{14}=(10.0,4.0)$,
$\x_4^{15}=(6.0,10.0)$, $\x_4^{16}=(0.0,6.0)$ activated at
$\tau_{4,1}=900$ps.  This MSS arrangement was obtained by seeking to
optimize the required simulation time, which, considering the
exponential decay of the photon density wave, was achieved by using
simultaneous laser sources that are as far away from each other as
possible---and thus facilitate the discrimination of signals received
at any given detector. In order to provide a sufficiently long
relaxation time for the photon density wave produced by the last
sources to be activated, each forward simulation for the MSS method
was carried on up to a final time $t_{\text{max}}=1400$ps. Similar
arrangements, whose details are not provided explicitly for the sake
of brevity, where used for the MSS cases $N_b=4$ and $N_b=8$. For the
TS method, in turn, each independent forward simulation was carried up
to the final time $t_{\text{max}}=600$ps, with detectors and sources
placed at the same position as for the single sources of the MSS
method with the corresponding value of $N_b$ and using the same
numerical grids, and number of processors for each case.

Figure~\ref{fig:itdet} shows that the TS and the MSS methods enjoy
similar convergence properties for a varying number $N_b$ of sources
and $N_d$ of detectors (cf. equation~\eqref{eq:FObj}). However, for
the $N_b=16$ benchmarks, the MSS method required $12,689$ seconds to
reach one-hundred iterations of the lm-BFGS algorithm while the TS
method, required $88,093$ seconds---so that the MSS method produces an
acceleration by a factor of almost seven to achieve one-hundred
lm-BFGS iterations. The figure also shows that increasing the number
of sources and detectors produces a significant improvement on the
convergence of the inverse solver for a fixed number of
iterations. The number of sources has a more significant impact on the
reconstructions than the number of detectors employed.  This can be
understood as follows: the role of the sources is to produce the
photon density waves that are used to sense the medium. Although the
detector reading mismatches $(G_j[u_i]-\widetilde{G}_{j,i})$ are
related to the sources for the adjoint
problem~\eqref{eq:AdjointProblem}, the intensity of these adjoint
sources depends, in turn, on the amount of photons reaching a given
detector.  It can be argued that for the same reason the number of
sources utilized has a more significant impact than whether the
sources are run on independent forward simulations (as in the TS
method), or simultaneously (as in the MSS method), which makes MSS
strategy a reliable and efficient approach.

Figure~\ref{fig:reconst} displays the true absorption coefficient as
well as the reconstructed absorption coefficients obtained by both the
TS and the MSS methods, with 16 sources and 36 detectors, after 100
iterations of the lm-BFGS algorithm. Clearly, the reconstructions
obtained by both methods are of comparable quality, but the MSS reconstruction resulted in a reduction in computing time by a factor of 7.

\subsection{Brain imaging based on hemodynamic activation or cancerous
  tissue}
In this final test case we consider a ``head model'' similar to the
one utilized in references~\cite{Prieto2017,Klose2002}.  This head
model mimics the typical situation where optical tomography is used to
study hemodynamic activity in the brain, and captures one of its
salient features, namely the clear layer that surrounds the
brain. This is a region between the scalp and the brain filled by
cerebrospinal fluid with negligible absorption and scattering
coefficients---which, in the context of optical tomography, is not
suitable for modeling under the diffusion
approximation~\cite{Boas2001} and thus requires use of the full RTE.
\begin{figure}[h!]
\centering
  \includegraphics[width=0.8\linewidth]{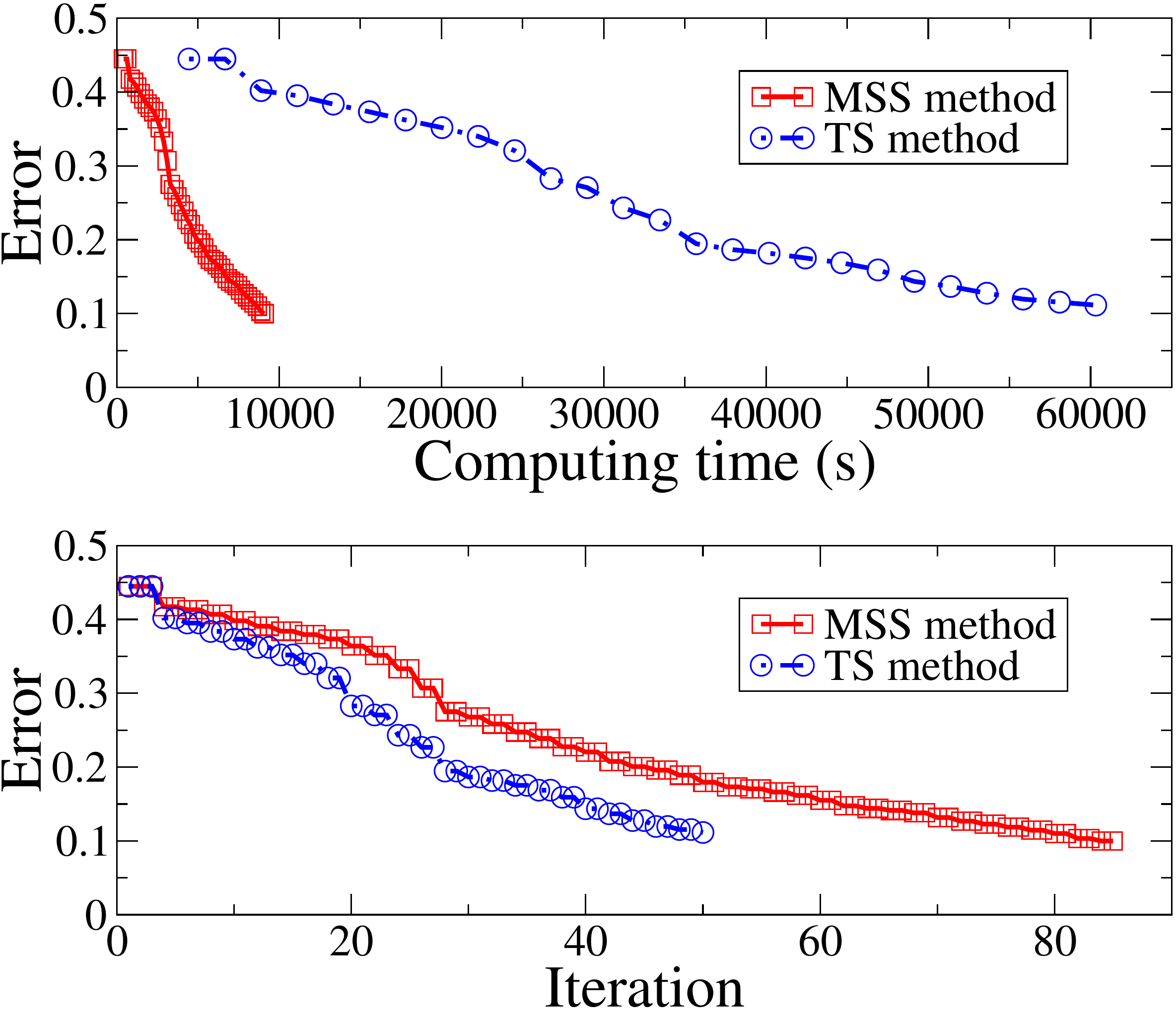}\\
  \caption{$L^2$-error norm eq.~\eqref{eq:errl2} evolution per iteration for the absorption coefficient 
  reconstruction in the head model for the TS and the MSS method.}
 \label{fig:ithead}
\end{figure}
\begin{figure*}[h!]
\centering
  \includegraphics[width=0.325\linewidth]{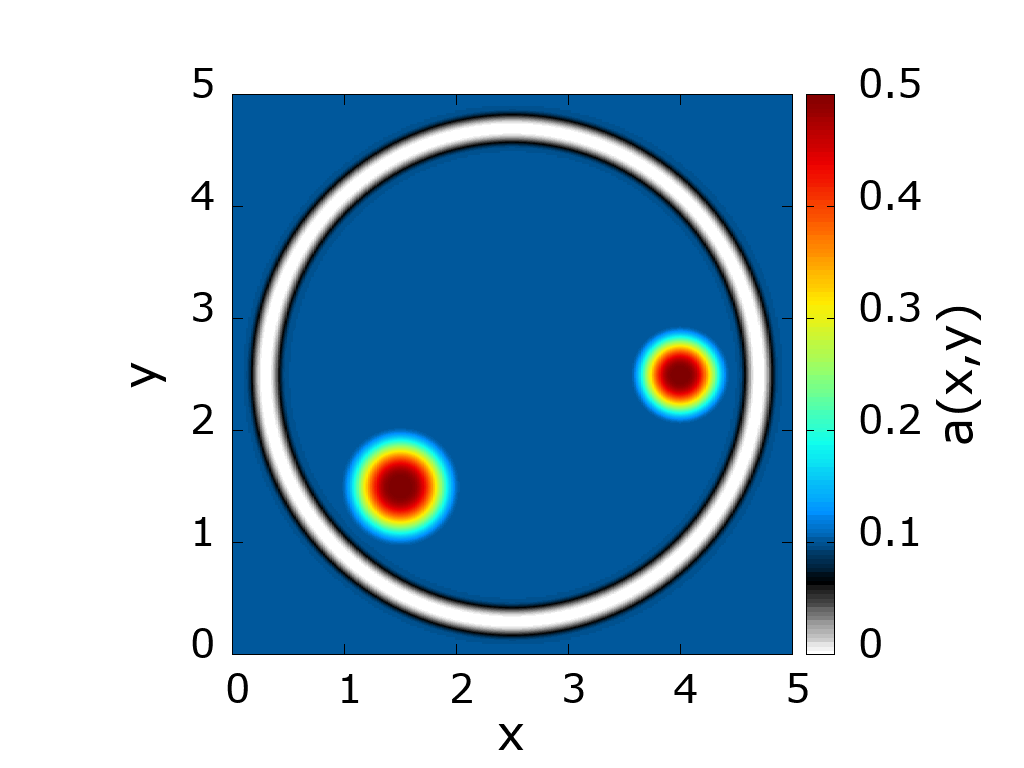}
  \includegraphics[width=0.325\linewidth]{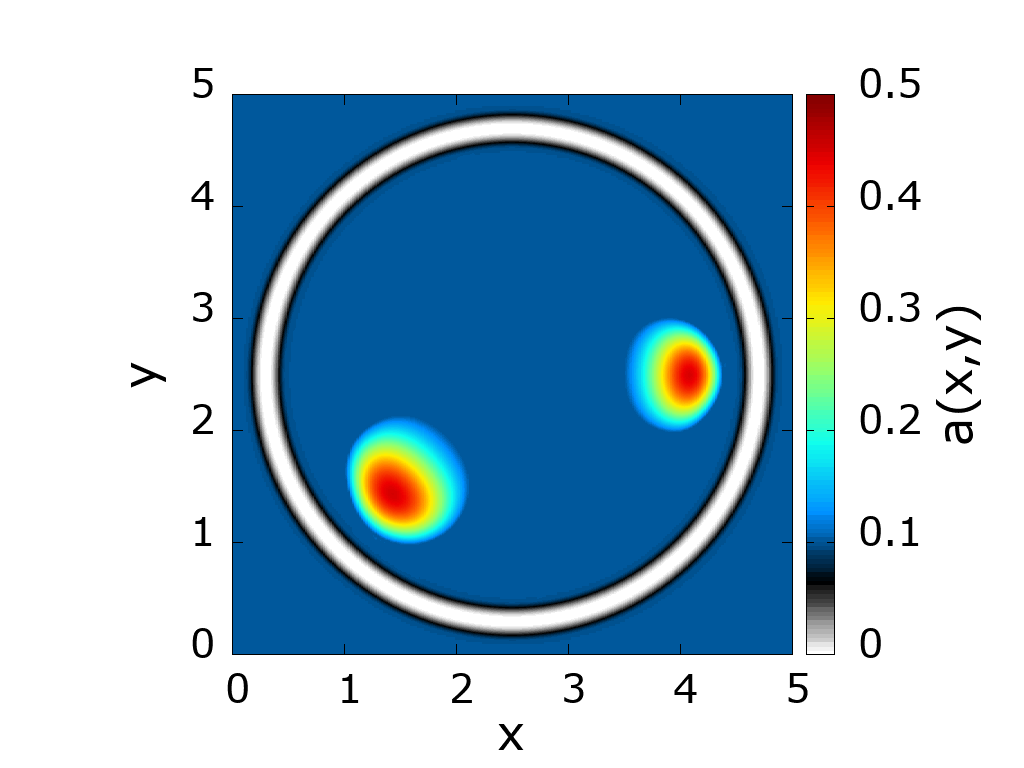}
  \includegraphics[width=0.325\linewidth]{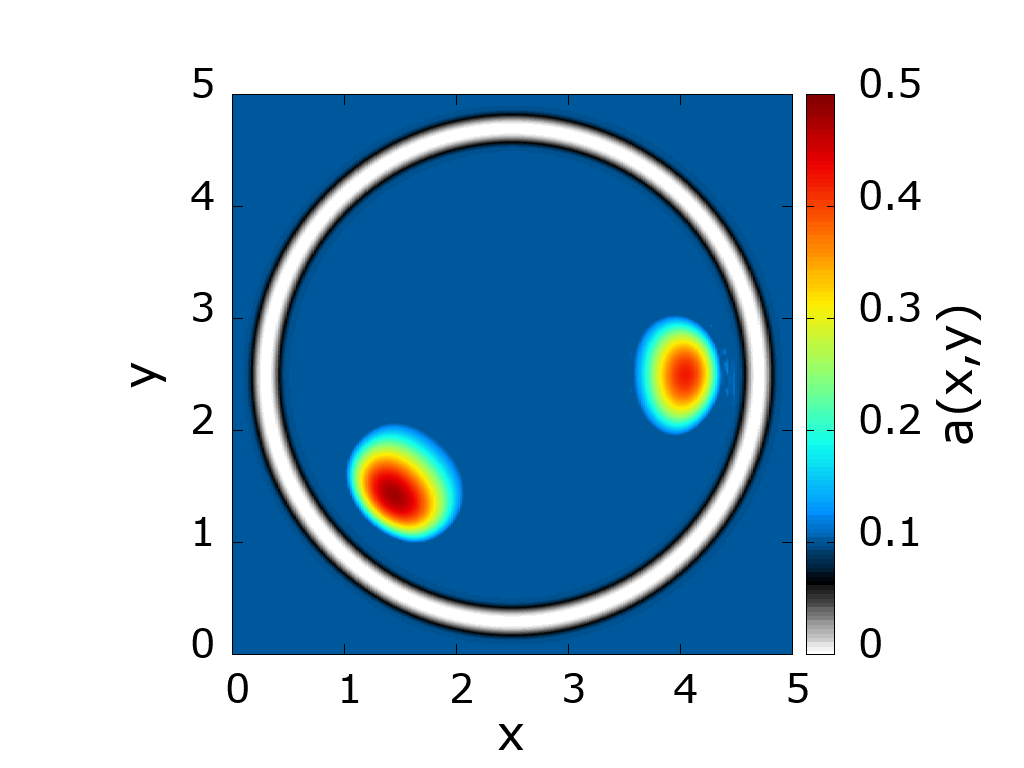}
  \caption{(Color online). From left to right: true absorption
    coefficient, reconstructed absorption coefficient after 50
    iterations of the TS method, and reconstructed absorption
    coefficient after 79 iterations of the MSS method. For these
    reconstructions 32 detectors and 16 sources were used. }
 \label{fig:rechead}
\end{figure*}

In the present context we employ the same number of sources as in
reference~\cite{Prieto2017}, where a total of 16 sources, with
four sources per face, are utilized for a similar head model. For our
reconstructions we again look for inclusions over a known background,
where the background value is used as the initial guess. We employ 32
detectors, with 8 detectors equally distributed per face. For this
benchmark all sources that are placed at the same face of the domain
are activated simultaneously.  By employing the notation previously
described in Section~\ref{sec:neck}, the activation configuration for
the MSS method for this benchmark is as follows: $\x_1^1=(1.0,0.0)$,
$\x_1^2=(2.0,0.0)$, $\x_1^3=(3.0,0.0)$, $\x_1^4=(4.0,0.0)$, activated
at $\tau_{1,1}=0$ps; $\x_2^5=(1.0,5.0)$, $\x_2^6=(2.0,5.0)$,
$\x_2^7=(3.0,5.0)$, $\x_2^8=(4.0,5.0)$ activated at
$\tau_{2,1}=100$ps; $\x_3^9=(5.0,1.0)$, $\x_3^{10}=(5.0,2.0)$,
$\x_3^{11}=(5.0,3.0)$, $\x_3^{12}=(5.0,4.0)$ activated at
$\tau_{3,1}=400$ps; and $\x_4^{13}=(0.0,1.0)$, $\x_4^{14}=(0.0,2.0)$,
$\x_4^{15}=(0.0,3.0)$, $\x_4^{16}=(0.0,4.0)$ activated at
$\tau_{4,1}=500$ps.  Figure~\ref{fig:ithead} displays the evolution of
the error eq.~\eqref{eq:errl2} for the TS and MSS methods, and
Figure~\ref{fig:rechead} presents the true absorption coefficient and
the final reconstructed absorption coefficients for the head model.

A reconstruction using the TS method for a total of 50 iterations is
presented in the middle panel of Figure~\ref{fig:rechead}. The right
panel in this figure, in turn, presents the results produced by MSS
using 79 iterations, which result in a similar error in the $L^2$
norm~\eqref{eq:errl2} as the one obtained from the 50 TS iterations.
The absorption in the clear layer and the region exterior to the clear
layer are assumed to be known and remain fixed during the
reconstruction process. The search for the inclusions is performed in
the region which is surrounded by the clear layer, given that this
would be the region where brain activity should be looked for in a
real optical tomography experiment. In terms of computational time,
the fifty iterations of the TS method required more than six times
longer than the 79 iterations required by the MSS method to achieve
the same error.


\section{Conclusions}
\label{sec:summary}

In this work we have considered the inverse RTE problem in optical
tomography by means of a non linear iterative optimization scheme
based on the lm-BFGS gradient descent method. We have demonstrated
that the proposed domain decomposition parallel strategy presents
perfect parallel scaling, making it suitable to tackle the
computational demands in optical tomography. Our newly obtained
adjoint-gradient expressions under Fresnel boundary conditions enables
correct accounting of refractive index mismatches at interfaces. In
conjunction with the MSS strategy, the algorithm provided a reduction
in computational times of several orders of magnitude over previous
approaches. This allowed us to produce reconstructions of the
absorption parameter in only a few hours, which in a single processor
and by means of the typically used TS method would have required
months of computing time to complete.

\section{Acknowledgements}
\label{sec:acknowledgements}
This work was supported by NSF, DARPA and AFOSR through contracts
DMS-2109831 and HR00111720035 and FA9550-21-1-0373, and
by the NSSEFF Vannevar Bush Fellowship under contract number
N00014-16-1-2808. ELG acknowledge financial support from CONICET.


\end{document}